\documentclass[twocolumn,showpacs,preprintnumbers,amsmath,amssymb,superscriptaddress]{revtex4}
\usepackage{epsfig}
\usepackage{psfrag}
\usepackage{amsfonts}
\usepackage{graphicx}
\usepackage{dcolumn}
\usepackage{bm}

\begin{document}


\title{Bose-Einstein Condensation of Bound Pairs of Relativistic Fermions in a Magnetic Field }

\author{Bo  Feng}
\affiliation{
School of Physics, Huazhong University of Science and Technology, Wuhan 430074, China
}
\affiliation{
Institute of Particle Physics and Key Laboratory of Quark and Lepton Physics
(MOE),  Central China Normal University, Wuhan 430079, China}

\author{De-fu Hou}
\affiliation{
Institute of Particle Physics and Key Laboratory of Quark and Lepton Physics
(MOE),  Central China Normal University, Wuhan 430079, China}

\author{Hai-cang Ren}
\affiliation{
Physics Department, The Rockefeller University, 1230 York Avenue, New York, New York 10021-6399, USA}
\affiliation{
Institute of Particle Physics and Key Laboratory of Quark and Lepton Physics
(MOE),  Central China Normal University, Wuhan 430079, China}
\author{Ping-ping Wu}
\affiliation{
School of Physics and Chemistry, Henan Polytechnic University, Jiaozuo 454000, China}

\date{\today}

\begin{abstract}
The Bose-Einstein condensation of bound pairs made of equally and oppositely charged  fermions in a magnetic field is investigated using a relativistic model.
The Gaussian fluctuations have been taken into account in order to study the spectrum of bound pairs in the strong coupling region. We found, in weak coupling reagion, the condensation temperature increases with an increasing magnetic field displaying the magnetic catalysis effect. In strong coupling region, the inverse magnetic catalysis appears when the magnetic field is low and is replaced by the usual magnetic catalysis effect when magnetic field is sufficiently high, in contrast to the nonrelativistic case where the inverse magnetic catalysis prevails in strong coupling region regardless of the strength of the magnetic field. The resulting
response to the magnetic field is the consequence of the competition between the dimensional reduction by Landau orbitals in
pairing dynamics and the anisotropy of the kinetic spectrum of the bound pairs. We thus conclude that dimensional reduction dominates in weak domain and strong coupling one except the small magnetic field region, where the enhanced fluctuations dominates.
\end{abstract}

\maketitle


\section{Introduction}

The phase structure of quantum chromodynamics(QCD) in the presence of an external magnetic field has been explored extensively in recent years\cite{lectnotes}.
This is of great phenomenological relevance for non-central heavy ion collisions and dense neutron stars as well as cosmological evolutions in its early stage.
In particular, the  peculiar response of the spontaneously breaking of chiral symmetry to a strong magnetic field is among the central issues and posseses a theoretical
challenge.

The profound effect of the magnetic field on chiral symmetry breaking had been pointed out  in quantum electrodynamics(QED) very early\cite{Klimenko,Miransky}.
It was found that magnetic fields can enhance the chiral symmetry breaking due to an effective dimension reduction in the lowest Landau level(LLL) in the charged
fermionic sector. This phenomenon is called \textit{magnetic catalysis}. In QCD, this problem can be addressed reliably by the lattice Monte-Carlo simulation without
sign problem\cite{latticetemp}. It has shown intriguingly and, however, unexpectedly that the chiral critical temperature decreases with an increasing magnetic field,
which is in explicit conflict with magnetic catalysis and thus termed as \textit{inverse magnetic catalysis} or \textit{magnetic inhibition}. Many attempts already exist
in the literatures trying to explain this contradiction\cite{Schmitt,Fukushimapawlowski, Fukushimahidaka,Kojo,Bruckmann, Mei,Pinto,Feng,Efrain,Paw,Braun,Mamo,Flachi,XGHuang,Kojo2}.

Along with the dimension reduction in the fermionic sector, which accounts for the magnetic catalysis, the significance of magnetic fields manifests in another important
aspect in a general way in systems with spontanously symmetry breaking. It enhances the fluctuations through the spatial anisotropy of the spectrum of bosonic modes in
the system, although the bosons are still in $3+1$ dimension and do not directly suffer from the Mermin-Wagner-Coleman theorem which forbids the spontaneously breaking of
continuous symmetry in $2+1$ and $1+1$ dimensions\cite{CMWH}. This claim follows by the observation that the Ginzburg critical window of the chiral phase transition gets
widened in the presence of a magnetic field\cite{Ginzburgwindow}. The role played by magnetic fields is thus twofold which competes with each other. A simple and straightforward model to
investigate this competition is the Bose-Einstein condensation(BEC) of neutral bosons composed by two equally and oppositely charged fermions\cite{Feng}, which is a
physical analog of chiral condensate in a magnetic field, yet analytically tractable. The corresponding non-relativistic case had been studied in \cite{Feng}, where we found that the critical temperature for the BEC
was dramatically affected by the magnetic field. In weak coupling domain (where the coupling fails to support a two fermion bound state in the absence of a magnetic field),
dimension reduction dominates and we found the usual magnetic catalysis effect. In strong
coupling domain (where a two fermion bound state exists in the absence of a magnetic field), however, the fluctuations dominate and the situation gets reversed, we found the inverse magnetic catalysis, i.e., the critical temperature of BEC decreases
as an increasing magnetic field.

In the present paper, we shall extend the above analysis to the relativistic fermions. Technically, this amounts to replace the non-relativistic fermionic propagator by the
Dirac one in a magnetic field\cite{Diracpropa}. The corresponding one loop diagram underlying the theory will be quadratically divergent. The leading divergence can be
removed by the coupling renormalization, similar to the non-relativistic case, where the renormalized coupling is in terms of the scattering length, but
the logarithmic subleading divergence remains and a UV cutoff has to be introduced by hand. Consequently, the results will be explicitly cutoff dependent, like all
relativistic field theories with 4-fermion interactions. Nevertheless, the cutoff does not change the results qualitatively.
We found that the low lying spectrum of the bound pairs is given by
\begin{equation}
\omega=\frac{p_3^2+\kappa {\bf p}_\perp^2}{4\bar{\mu}},
\end{equation}
where $p_3$($p_\perp$) is the momentum component parallel (perpendicular) to the magnetic field, $\kappa\le 1$ is an anisotropicity factor and $2\bar\mu$ is the
rest energy of a bound pair. This spectrum implies
a nonrelativistic like condensation temperature
\begin{equation}
T_C=\left[\frac{\kappa n}{2\zeta(3/2)}\right]^{2/3}\frac{\pi}{\bar\mu}\label{transitiontemperature}.
\end{equation}
at sufficiently low density.
Unlike the nonrelativistic case where only the anisotropicity factor $\kappa$ is magnetic field dependent, the "mass" $\bar\mu$ here depends on the magnetic field
as well. In the weak coupling domain, the magnetic field dependence of both $\kappa$ and $\bar\mu$ support magnetic catalysis and $T_C$ increases with an increasing
magnetic field as in the nonrelativistic case.
The situation in strong coupling domain, however, is very different from that in the nonrelativistic case,
in which the inverse magnetic catalysis prevails regardless of the strength of the magnetic field. In the present case, the inverse magnetic catalysis only appears
when magnetic field is low and the magnetic catalysis effect takes place again when magnetic field is sufficiently high because of opposite roles played
by $\kappa$ and $\bar\mu$ there. Consequently, $T_C$ decreases first and then increases with the magnetic field.

The rest of the paper is organized as follows: in Section II we lay out the general formulation and present the mean field
approximation.  The fluctuations beyond the mean field theory, which is necessary for BEC, is calculated
under the Gaussain approximation in Section III. The numerical solutions to the gap equation and the condensation temperature are presented in Section IV. Section V is devoted to the conclusions and outlooks. Some calculation details and
useful formulas are presented in the Appendices A, B, C and D. Throughout the paper, we will work in Euclidean space with the
four vector represented by $x^\mu=(i\tau,{\bf x}), {q^\mu=(i\omega_n, {\bf q})}$ with $\omega_n$ the Matsubara frequency for
bosons $\omega_n=2i\pi nT$ or for fermions $\omega_n=(2n+1)i\pi T$ with $T$ the temperature.

\section{The Model and Its Mean field Approximation}

We consider a system consisting of relativistic fermions of mass $m$ and chemical potential $\mu$ with opposite electric charge
interacting through a short ranged instantaneous attractive interaction. In a quark matter, the attractive interaction stems from
the nonperturbative QCD effects and the long range Coulomb interaction is perturbative and can be ignored in the leading order.
The Lagrangian density reads
\begin{equation}
{\cal L}={\cal L}_0+{\cal L}_I,
\end{equation}
where
\begin{equation}
{\cal L}_0=\sum_{\alpha=\pm}{\bar{\psi}_\alpha}\left(i\gamma^\mu D_\mu-m+\mu\gamma^0\right)\psi_\alpha,
\end{equation}
and
\begin{equation}
{\cal L}_I=G\bar{\psi}_+ i\gamma^5\psi_-^C\bar{\psi}_-^C i\gamma^5\psi_+.
\end{equation}
with $D_\mu=\partial_\mu-i\alpha e A_\mu$ the covariant derivative  and $e>0$ the charge magnitude. We also defined the charge conjugate operator $C=i\gamma^2\gamma^0$ and $\psi^C=C\bar{\psi}^T, \bar{\psi}^C=\psi^T C, \psi=C(\bar{\psi}^C)^T, \bar{\psi}=(\psi^C)^T C$. In the following, we shall consider the situation of a constant magnetic field and work in Landau gauge, in which the vector potential $A_x=A_z=0, A_y=Bx$(the field $B$ being along the $z$ direction).

Introducing the usual Hubbard-Stratonovich field $\Delta(x)$ coupled to $\bar{\psi}_+ i\gamma^5\psi_-^C$, we obtain the partition function
\begin{equation}
{\cal Z}={\cal N}\int{\cal D}\bar{\Psi}(x){\cal D}\Psi(x){\cal D}\Delta(x){\cal D}\Delta^*(x)\exp\left[{\cal S}_{\rm eff.}\right].
\end{equation}
where $\cal N$ is the normalization constant and the Nambu-Gorkov(NG) spinors are defined as
\begin{equation}
\Psi(x)=\left(\begin{array}{c}
\psi_+(x)\\
\psi_-^C(x)\\
\end{array}\right),\hspace{0.5cm}
\bar{\Psi}(x)=\left(\begin{array}{cc}
\bar{\psi}_+(x), &
\bar{\psi}_-^C(x)\\
\end{array}\right),
\end{equation}
The effective action  is given by
\begin{equation}
{\cal S}_{\rm eff.}=\int d^4 x\left[\int d^4 y\bar{\Psi}(x)G^{-1}(x,y)\Psi(y)-\frac{|\Delta(x)|^2}{G}\right],
\end{equation}
with
\begin{widetext}
\begin{equation}
G^{-1}(x,y)=\left(\begin{array}{cc}
-\gamma^0\frac{\partial}{\partial\tau}+\gamma\cdot(i\nabla+e{\bf A})+\mu\gamma^0-m & i\gamma^5\Delta(x)\\
i\gamma^5\Delta^*(x) & -\gamma^0\frac{\partial}{\partial\tau}+\gamma\cdot(i\nabla+e{\bf A})-\mu\gamma^0-m\\
\end{array}\right)\delta^4(x-y).\label{NGpropagator}
\end{equation}
\end{widetext}
The thermodynamic potential reads
\begin{equation}
\Omega=-T\ln{\cal Z},
\end{equation}
and the number density of fermions is given by
\begin{equation}
n=\frac{1}{V}\left(\frac{\partial P}{\partial\mu}\right)_T.
\label{fermionnumberdensity}
\end{equation}
with $P=-\Omega$ the pressure and $V$ the spatial volume of the system.
Integrating out the NG degrees of freedom, the partition function becomes
\begin{equation}
{\cal Z}={\cal N}\int{\cal D}\Delta(x){\cal D}\Delta^*(x)\exp\left({\cal S_{\rm eff.}}[\Delta(x)]\right),
\end{equation}
with
\begin{equation}
{\cal S}_{\rm eff.}={\rm Tr}\ln G^{-1}[\Delta(x)]-\int d^4 x\frac{|\Delta(x)|^2}{G}.\label{effectiveaction}
\end{equation}
where the trace runs over space, imaginary time, Dirac and NG indices.

The mean field approximation ignores the fluctuations in $\Delta(x)$ field, which amounts to set $\Delta(x)=\Delta_0$. In this approximation, we obtain
\begin{align}
\nonumber & {\cal S}_{\rm eff.}=-\frac{\beta V}{G}\Delta_0^2+\beta VT\sum_{n}\int\frac{dp_2dp_3}{(2\pi)^2}\sum_{l=0}^{\infty}\alpha_l\\
&\times\ln\left[\left((i\omega_n)^2-\left[E_l^+(\Delta_0)\right]^2\right)
\left((i\omega_n)^2-\left[E_l^-(\Delta_0)\right]^2\right)\right],
\end{align}
where $\beta=1/T$ and $l=0,1,2,\cdots$ labels the Landau levels (LL).
Notice that, the factor $\alpha_l=2-\delta_{l,0}$ reflects the non-degeneracy in the Lowest Landau Level (LLL). We defined
\begin{equation}
E_l^\pm(\Delta_0)=\left[\left(\sqrt{p_3^2+2leB+m^2}\pm\mu\right)^2+\Delta_0^2\right]^{1/2}.
\end{equation}
which are the energy spectrum of the fermionic exitations in the system. The thermodynamic potential under mean-field approximation is
\begin{equation}
\Omega = -T{\cal S}_{\rm eff.}[\Delta_0].
\label{pressure0}
\end{equation}

From the saddle point condition
\begin{equation}
\frac{\delta{\cal S}_{\rm eff.}[\Delta_0]}{\delta\Delta_0}=0,
\end{equation}
one obtains the gap equation
\begin{align}
\nonumber\frac{1}{G}=&\frac{eB}{(2\pi)^2}\int_0^\infty dp_3\sum_{l=0}^{\infty}\alpha_l\left[\frac{1}{E_l^+(\Delta_0)}\tanh\left(\frac{E_l^+(\Delta_0)}{2T}\right)\right.\\
&+\left.\frac{1}{E_l^-(\Delta_0)}\tanh\left(\frac{E_l^-(\Delta_0)}{2T}\right)\right].\label{gapequation}
\end{align}
and $\Delta_0=0$ at the transition temperatute $T_c$. The number density under mean field approximation follows from eq.(\ref{fermionnumberdensity}) and reads
\begin{equation}
n=-\frac{T}{V}\left(\frac{{\cal S}_{\rm eff.}[\Delta_0]}{\partial\mu}\right)_T.
\label{fermionnumberdensity0}
\end{equation}

In BCS limit, the critical temperature coincides with the threshold pairing temperature which, at a given chemical potential, is determined by (\ref{gapequation}) at $\Delta_0=0$
first and the chemical potential is in turn solved by the density equation (\ref{fermionnumberdensity}).
In BEC limit, however, the roles of (\ref{gapequation}) and (\ref{fermionnumberdensity}) reversed because the critical tempearture becomes separate from the
pairing. Eq. (\ref{gapequation}) determines the chemical potential\cite{Sademelo}.
The BEC critical temperature is determined by  eq.(\ref{fermionnumberdensity}) equals (\ref{fermionnumberdensity0}) plus the contribution from the bosonic excitation
spectrum, whose low momentum behavior will be given in the next section. There we shall also see that the quantity $2(m-\mu)>0$ with $\mu$ the solution of the gap
equation measures the binding energy of a relativistic pair and locates the bosonic pole below the two-fermion cut at zero total momentum.

To make the formulation analytically tractable, we assume that
$-m<\mu<m$ and $T<<m\pm\mu$ the gap equation (\ref{gapequation}) can be approximated by
\begin{align}
\nonumber\frac{1}{G}&=\frac{eB}{(2\pi)^2}\int_0^\infty dp_3\sum_{l=0}^{\infty}\alpha_l\\
&\times\left[\frac{1}{\sqrt{p_3^2+2leB+m^2}+\mu}+\frac{1}{\sqrt{p_3^2+2leB+m^2}-\mu}\right].\label{zerotemgapeq}
\end{align}
where we have set $\Delta_0=0$, which amounts to approach the critical point from normal to superconducting phase. Obviously, this equation is quadratically divergent and needs to be regularized.
To maintain the relativistic invariance, we shall use the Pauli-Villars regularization scheme. Introducing the PV regulators, we obtain
\begin{align}
\nonumber\frac{1}{G}&=\frac{eB}{(2\pi)^2}\sum_s C_s\int_0^\infty dp_3\sum_{l=0}^{\infty}\alpha_l\\
&\times\left[\frac{1}{\sqrt{p_3^2+2leB+M_s^2}+\mu}+\frac{1}{\sqrt{p_3^2+2leB+M_s^2}-\mu}\right].\label{PVgapeq}
\end{align}
with $M_0^2=m^2, C_0=1$ and $\sum_s C_s=0, \sum_s C_s M_s^2=0$. Throughout the paper, we shall assume that the regulator mass $M_{s>0}>>m$ but finite.
Carrying out the integral over the momentum under the constrains imposed by the regularization conditions, we obtain
\begin{align}
\nonumber\frac{1}{G}=&-\frac{eB}{(2\pi)^2}\sum_{l=0}^{\infty}\alpha_l\sum_s C_s\Big[\ln(2leB+M_s^2)\\
&-\frac{2\mu}{\sqrt{2leB+M_s^2-\mu^2}}\tan^{-1}\frac{\mu}{\sqrt{2leB+M_s^2-\mu^2}}\Big].\label{gapeqinappendix}
\end{align}
Dropping all terms with negative powers of $M_{s>0}$ since $m/M_s<<1$ by definition, we end up with
\begin{widetext}
\begin{align}
\nonumber\frac{1}{G}=&\frac{eB}{2\pi^2}\left\{\left[\ln\Gamma\left(\frac{m^2}{2eB}\right)-\frac{\mu^2}{eB}\psi\left(\frac{m^2-\mu^2}{2eB}\right)-\frac{\mu}{\sqrt{m^2-\mu^2}}\tan^{-1}\frac{\mu}{\sqrt{m^2-\mu^2}}\right]\right.\\
\nonumber&+\left.2\sum_{l=0}^{\infty}\left[\frac{\mu}{\sqrt{2leB+m^2-\mu^2}}\tan^{-1}\frac{\mu}{\sqrt{2leB+m^2-\mu^2}}-\frac{\mu^2}{2leB+m^2-\mu^2}\right]\right\}\\
&+\frac{1}{4\pi^2}\left[m^2\sum_{s>0}C_s\Lambda_s^2\ln\Lambda_s^2-2\mu^2\sum_{s>0}C_s\ln\Lambda_s^2+m^2\left(1+\ln\frac{2eB}{m^2}\right)-2\mu^2\ln\frac{2eB}{m^2}-eB\ln\frac{4\pi eB}{m^2}\right].\label{regularizedgapeq}
\end{align}
\end{widetext}
with $\Lambda_s^2\equiv\frac{M_s^2}{m^2}$ and $\Gamma(x)$ the gamma function, and $\psi(x)=d\ln\Gamma(x)/dx$.
The detailed steps leading from (\ref{gapeqinappendix}) to (\ref{regularizedgapeq}) will be shown in Appendix A.
The quadratic term in the regulator masses in the last line can be absorbed into the coupling constant as usual and the logarithmic term in the regulator ones remains, leaving the gap equation explicitly depends on an UV cutoff. In the appendix B, we have proved that solutions to this gap equation is a decreasing function of an increasing magnetic field.

The gap equation at zero magnetic field can be obtained straightforwardly from
(\ref{regularizedgapeq}) and reads
\begin{align}
\nonumber\frac{1}{G_R}=&-\frac{\mu^2}{2\pi^2}\sum_{s>0}C_s\ln\Lambda_s^2\\
&+\frac{\mu^2}{\pi^2}-\frac{\mu\sqrt{m^2-\mu^2}}{\pi^2}\tan^{-1}\frac{\mu}{\sqrt{m^2-\mu^2}}.\label{gapeqatzerofield}
\end{align}
where the critical coupling $G_R$ had absorbed the quadratically divergent term. The strong and weak coupling domain is characterized by a critical coupling, which is the threshold
coupling for a bound state to show up in the absence of a magnetic field. It follows then the critical coupling constant is defined as the one for $\mu=m$. Moreover, we shall denote $\sum_sC_s\ln\Lambda_s=-\ln\Lambda\equiv\ln\frac{M}{m}$ with $M$ an UV cutoff.  We thus have
\begin{equation}
\frac{1}{G_R^Cm^2}=\frac{1}{2\pi^2}\ln\Lambda^2+\frac{1}{\pi^2}.\label{criticalcoupling}
\end{equation}
This definition establishs an explicit dependence of the dimensionless critical coupling constant $G_R^Cm^2$ on the UV cutoff.

\section{The Gaussian Fluctuations}

In this section, we shall include all nonzero momentum component of the Hubbard-Stratonovich field to the quadratic order (Gaussian fluctuations) in the partition function.
Approching the condensate from the normal phase, we shall reproduce the gap quation (\ref{zerotemgapeq}) by the Thouless critirion and obtain the contribition from
the bosonic excitations to the number density (\ref{fermionnumberdensity}), which is necessary to determine the critical temperature in BEC limit. For the simplicity of notation, we shall not explicitly write out the Pauli-Villars regulators in the following calculations. But they will be restored whenever the integral is divergent.

Expanding the bosonic field $\Delta(x)$ around the trivial saddle point $\Delta_0=0$ in the effective action (\ref{effectiveaction}), we obtain that
\begin{align}
\nonumber{\cal S}_{\rm eff}[\Delta(x)]&\simeq {\cal S}_{\rm eff}[0]-\frac{1}{G}\int d^4 x|\Delta(x)|^2\\
&+{\rm Tr}\left[G_0^-(x,y)\gamma^5\Delta^*(y) G_0^+(y,x)\gamma^5 \Delta(x)\right],\label{guanssianaction}
\end{align}
with $[G_0^\pm]^{-1}$ the diagonal elements in the NG propagator (\ref{NGpropagator}), which are nothing but the inverse Dirac operators in a magnetic field in coordinate space. The Dirac propagator had been firstly obtained by Schwinger using the proper time method.
Here we shall employ the form with explicit Landau levels dependence, given in\cite{Diracpropa}
\begin{align}
\nonumber G_0^\pm(x,y)\equiv&\left[-\gamma^0\left(\partial_\tau\pm\mu\right)+i\gamma\cdot {\bf D}-m\right]^{-1}\delta(x-y)\\
=& e^{i\Phi(x,y)}T\sum_n\int\frac{d^3{\bf k}}{(2\pi)^3}e^{-ik\cdot z}\sum_{l=0}^{\infty}G_l^\pm(\omega_n,{\bf k}),\label{Diracpropagator}
\end{align}
with
\begin{widetext}
\begin{align}
\nonumber G_l^\pm(\omega_n,{\bf k})
\nonumber=& e^{\frac{-{\bf k}_\perp^2}{eB}}\frac{(-1)^l}{(i\omega_n\pm\mu)^2-k_3^2-2leB-m^2}\left\{\left[\gamma^0(i\omega_n\pm\mu)-\gamma^3k^3+m\right]\left[2\Pi_+L_l\left(\frac{2{\bf k}_\perp^2}{eB}\right)-2\Pi_-L_{l-1}\left(\frac{2{\bf k}_\perp^2}{eB}\right)\right]\right.\\
&+\left.4({\bf k}\cdot\gamma)_\perp L_{l-1}^1\left(\frac{2{\bf k}_\perp^2}{eB}\right) \right\}.
\end{align}
\end{widetext}
where $z=x-y, {\bf k}_\perp=(k_1,k_2)$ and we also defined the projectors $\Pi_\pm=(1\pm i\gamma^1\gamma^2)/2$, $L_l^\alpha(x)$ is the generalized Laguerre polynomials($L_l\equiv L_l^0$ and $L_{-1}^\alpha=0$ by definition). The gauge dependent phase factor is given by
\begin{equation}
\Phi(x,y)=-\frac{eB}{2}(x_1+y_1)(x_2-y_2).
\end{equation}
which cancels out in the combination of (\ref{guanssianaction}). Consequently, the kernel mediating $\Delta^*(y)$ and $\Delta(x)$ there is explictly
translational invariant, i.e., depends only on $x-y$, as expected.

In momentum space, eq. (\ref{guanssianaction}) becomes
\begin{equation}
{\cal S}_{\rm eff}[\Delta]={\cal S}_{\rm eff}[0]-T\sum_n\int \frac{{d^3\bf p}}{(2\pi)^3}\Gamma^{-1}(\omega_n,{\bf p})|\Delta(\omega_n,{\bf p})|^2,\label{gaussianeffectiveaction}
\end{equation}
where the dependence of the coefficient $\Gamma^{-1}$ on $T,\mu$ and $B$ is implicit. The instability towards the superconducting phase is then signaled by the condition
\begin{equation}
\Gamma^{-1}(0,0)=0.\label{thoulesseq}
\end{equation}
which is just the Thouless criterion for the onset of a long range order. The number density equation is determined by
\begin{equation}
n=n_0-\frac{1}{\beta }\frac{\partial}{\partial\mu}T\sum_n\int \frac{{d^3\bf p}}{(2\pi)^3}\ln\Gamma(\omega_n, {\bf p}),\label{gaussiannumbereq}
\end{equation}
where $n_0=\sum_l\int dp_2dp_3/(2\pi)^2[n_F(E_l^+(0))-n_F(E_l^-(0))]$ with $n_F(\omega)=1/(\exp(\beta\omega)+1)$ the Fermi-Dirac distribution function,
is the fermionic contribution defined in (\ref{fermionnumberdensity}). The eqs. (\ref{thoulesseq}) and (\ref{gaussiannumbereq}) are two basic equations to solve for BCS/BEC crossover\cite{Sademelo,Nozieres,abuki,Dengjian,Efraincrossover,Zhuang}. Following the convention in\cite{Nozieres}, one can write the number density (\ref{gaussiannumbereq}) in terms of a phase shift defined by
$\Gamma(\omega\pm i0,{\bf q})=|\Gamma(\omega,{\bf q})|\exp[\pm i\delta(\omega, {\bf q})]$. We have
\begin{equation}
n=n_0+\int\frac{d^3{\bf q}}{(2\pi)^3}\int_{-\infty}^{\infty}\frac{d\omega}{\pi}n_B(\omega)
\frac{\partial\delta}{\partial\mu}(\omega,{\bf q}).\label{numberdensity}
\end{equation}
with $n_B(\omega)=1/[\exp(\beta\omega)-1]$ the Bose-Einstein distrubution function.

To calculate $\Gamma^{-1}$, we make use of the eq.(\ref{Diracpropagator}) and the Fourier transformation of the fluctuation field
\begin{equation}
\Delta(x)=T\sum_n\int\frac{d^3{\bf k}}{(2\pi)^3}e^{-ik\cdot x}\Delta(\omega_n,{\bf k}),
\end{equation}
and obtain
\begin{widetext}
\begin{align}
\nonumber\Gamma^{-1}(\omega_n,{\bf p})=&-8T\sum_{n^\prime}\int\frac{d^3{\bf k}}{(2\pi)^3}e^{-\frac{{\bf k}_\perp^2+{\bf q}_\perp^2}{eB}}\sum_{l,l^\prime}\frac{(-1)^{l^\prime}}{(i\omega_{n^\prime}-\mu)^2-k_3^2-2{l^\prime}eB-m^2}\frac{(-1)^{l}}{(i\omega_{n^\prime}+i\omega_n+\mu)^2-q_3^2-2leB-m^2}\\
\nonumber&\times\left\{ \left[(i\omega_{n^\prime}-\mu)(i\omega_{n^\prime}+i\omega_n+\mu)-k_3q_3-m^2\right]\left[L_{l^\prime}\left(\frac{2{\bf p}_\perp^2}{eB}\right)L_l\left(\frac{2{\bf q}_\perp^2}{eB}\right)+L_{l^\prime-1}\left(\frac{2{\bf p}_\perp^2}{eB}\right)L_{l-1}\left(\frac{2{\bf q}_\perp^2}{eB}\right)\right]\right.\\
&\left.-8\left({\bf k}\cdot{\bf q}\right)_\perp L^1_{l^\prime-1}\left(\frac{2{\bf p}_\perp^2}{eB}\right)L^1_{l-1}\left(\frac{2{\bf q}_\perp^2}{eB}\right)\right\}-\frac{1}{G}.
\end{align}
\end{widetext}
with ${\bf q}={ \bf k}+\bf p$.
Working out the sum over Matsubara frequecies, one can verify that the condition (\ref{thoulesseq}) will yield the same gap equation as (\ref{zerotemgapeq}). Let's proceed by employing the proper-time representation of the propagators introduced  by Schwinger
\begin{equation}
\int_0^1 dx\int_0^\infty d\lambda \lambda e^{\lambda( xD_1+(1-x)D_2)}=\frac{1}{D_1D_2}.
\end{equation}
where the Feynmann parametrization scheme had been used. We obtain
\begin{widetext}
\begin{align}
\nonumber\Gamma^{-1}(\omega_n,{\bf p})=& 8i\int\frac{d\omega'd^3{\bf k}}{(2\pi)^4}e^{-\frac{{\bf k}_\perp^2+{\bf q}_\perp^2}{eB}}\sum_{l,l^\prime}(-1)^{l+l^\prime}\int_0^1 dx\int_0^\infty d\lambda \lambda e^{\lambda( x\left[\omega^{{\prime}2}-k_3^2-2{l^\prime}eB-m^2 \right]+(1-x)\left[(\omega'+\omega+2\mu)^2-q_3^2-2leB-m^2 \right] )}
\\
\nonumber&\times\left\{ \left[\omega^{\prime}(\omega^{\prime}+\omega+2\mu)-k_3q_3-m^2\right]\left[L_{l^\prime}\left(\frac{2{\bf p}_\perp^2}{eB}\right)L_l\left(\frac{2{\bf q}_\perp^2}{eB}\right)+L_{l^\prime-1}\left(\frac{2{\bf p}_\perp^2}{eB}\right)L_{l-1}\left(\frac{2{\bf q}_\perp^2}{eB}\right)\right]\right.\\
&\left.-8\left({\bf k}\cdot{\bf q}\right)_\perp L^1_{l^\prime-1}\left(\frac{2{\bf p}_\perp^2}{eB}\right)L^1_{l-1}\left(\frac{2{\bf q}_\perp^2}{eB}\right)\right\}-\frac{1}{G}.\label{inversegamma}
\end{align}
\end{widetext}
where we analytically continuated the Matsubara frequency to the continuous one as $i\omega_n\rightarrow \omega+i\epsilon^+$ and shift the frequency from $\omega^\prime$ to $\omega^\prime+\mu$. The same approximation as in the preceding section, $T<<m-|\mu|$ has been made in (\ref{inversegamma}) that turns the Matsubara sum over $n'$ to the integral over $\omega'$

Using the generating function of Laguerre polynomials
\begin{equation}
\sum_{l=0}^\infty s^l L_l^n(z)=\frac{1}{(1-s)^{n+1}}e^{-\frac{zs}{1-s}},
\end{equation}
one can carry out the sum over Landau levels and has
\begin{widetext}
\begin{align}
\nonumber\Gamma^{-1}(\omega,{\bf p})= & 8i\int\frac{d^2kd^2{\bf k}_\perp}{(2\pi)^4}\int_0^1 dx\int_0^\infty d\lambda \lambda e^{-\frac{1+ t}{1-t}\frac{{\bf k}_\perp^2}{eB}-\frac{1+ t'}{1-t'}\frac{{\bf q}_\perp^2}{eB}+\lambda\left[x\left(k^2-m^2 \right)+(1-x)\left(q^2-m^2\right)\right]}\\
&\times \left\{ \left[k\cdot q-m^2\right]\frac{1+tt'}{(1-t)(1-t')}-8\left({\bf k}\cdot{\bf q}\right)_\perp\frac{tt'}{(1-t)^2(1-t')^2} \right\}-\frac{1}{G}.\label{covariantgamma}
\end{align}
\end{widetext}
with $t=-e^{-2\lambda xeB}$ and $t'=-e^{-2\lambda(1-x)eB}$. We defined $k\equiv (\omega',k_3)$, $p\equiv (\omega+2\mu,p_3)$ and $q\equiv k+p$, by which eq.(\ref{covariantgamma}) can be written in a covariant form. One may observe that all momentum integrals become Gaussian ones, which is the advantage of using the proper-time representations. The integral over $k$ can be carried out by shifting the variable to $k+(1-x)p$ and Wick rotating $k^0=ik_E^0, k^3=k_E^3$ with $k_E^0$ from $-\infty$ to $\infty$, and we obtain
\begin{widetext}
\begin{align}
\nonumber\Gamma^{-1}(\omega,{\bf p})= & \frac{1}{2\pi^3}\int d^2{\bf k}_\perp\int_0^1 dx\int_0^\infty d\lambda  e^{-\frac{1+ t}{1-t}\frac{{\bf k}_\perp^2}{eB}-\frac{1+ t'}{1-t'}\frac{{\bf q}_\perp^2}{eB}+\lambda\left[x(1-x)p^2-m^2\right]}\\
&\times \left\{ \left[\frac{1}{\lambda}+x(1-x)p^2+m^2\right]\frac{1+tt'}{(1-t)(1-t')}+8\left({\bf k}\cdot{\bf q}\right)_\perp\frac{tt'}{(1-t)^2(1-t')^2} \right\}-\frac{1}{G}.
\end{align}
\end{widetext}
The other integral over momentum ${\bf k}_\perp$ can be performed by noticing that
\begin{align}
\int\frac{d^2{\bf k}_\perp}{(2\pi)^2}e^{-y{\bf k}_\perp^2-y'{\bf q}_\perp^2}=\frac{1}{4\pi(y+y')}e^{-\frac{yy'}{y+y'}{\bf p}_\perp^2},
\end{align}
and
\begin{align}
\nonumber\int\frac{d^2{\bf k}_\perp}{(2\pi)^2}({\bf k}\cdot {\bf q})_\perp e^{-y{\bf k}_\perp^2-y'{\bf p}_\perp^2}=&\frac{1}{4\pi(y+y')^2}e^{-\frac{yy'}{y+y'}{\bf p}_\perp^2}\\
&\times\left(1-\frac{yy'}{y+y'}{\bf p}_\perp^2 \right).
\end{align}
we end up with
\begin{widetext}
\begin{align}
\nonumber \Gamma^{-1}(\omega,{\bf p})
= &\frac{eB}{4\pi^2}\int_0^1 dx\int_0^\infty d\lambda   e^{\lambda[x(1-x)p^2-m^2] -\frac{\tanh(\lambda xeB)\tanh(\lambda(1-x)eB)}{\tanh(\lambda xeB)+\tanh(\lambda(1-x)eB)}\frac{{\bf p}_\perp^2}{eB}}\left\{ \left[\frac{1}{\lambda}+x(1-x)p^2+m^2\right]\coth(\lambda eB)\right.\\
&+\left.\frac{1}{\sinh^2(\lambda eB)} \left[eB -\frac{\tanh(\lambda xeB)\tanh(\lambda(1-x)eB)}{\tanh(\lambda xeB)+\tanh(\lambda(1-x)eB)}{\bf p}_\perp^2\right] \right\}-\frac{1}{G}.\label{finalgamma}
\end{align}
\end{widetext}

The singularity structure of $\Gamma(\omega,{\bf p})$ in the entire complext $\omega$-plane
reflects the two-particle spectrum. There will be an isolated real pole representing the two-body bound pair and
a branch cut along the real axis representing the continuum of two-particle excitations. For sufficiently strong coupling and low temperature,
the contribution to the number density is dominated by the bound pair pole, which is determined by $\omega=0, {\bf p}=0, \mu={\bar\mu}$ with ${\bar\mu}$
the solution to the mean field equation (\ref{regularizedgapeq}). We henceforth consider the expansion of (\ref{finalgamma})
around this pole and obtain
\begin{equation}
\Gamma^{-1}(\omega,{\bf p})=\Gamma^{-1}(0,0)+a\left(\omega-\frac{p_3^2}{4{\bar{\mu}}}\right)-b\frac{{\bf p}_\perp^2}{4{\bar{\mu}}}+...,\label{gammaundergaussian}
\end{equation}
with
\begin{align}
\nonumber & a=\frac{eB{\bar{\mu}}}{\pi^2}\int_0^1 dx\int_0^\infty d\lambda \lambda x(1-x)e^{\lambda\left[4x(1-x)\bar{\mu}^2-m^2\right]}\\
&\times\left\{\left[\frac{2}{\lambda}+4x(1-x)\bar{\mu}^2+m^2\right]\coth(\lambda eB)+\frac{eB}{\sinh^2(\lambda eB)} \right\},
\label{coeffa}
\end{align}
and
\begin{align}
\nonumber & b=\frac{{\bar{\mu}}}{\pi^2}\int_0^1 dx\int_0^\infty d\lambda e^{\lambda\left[4x(1-x)\bar{\mu}^2-m^2\right]}\\
\nonumber&\times\frac{\tanh(\lambda xeB)\tanh(\lambda(1-x)eB)}{\tanh(\lambda xeB)+\tanh(\lambda(1-x)eB)}\\
&\times \left\{\left[\frac{1}{\lambda}+4x(1-x)\bar{\mu}^2+m^2\right]\coth(\lambda eB)+\frac{2eB}{\sinh^2(\lambda eB)}\right\}.
\label{coeffb}
\end{align}
Notice that,  the kinetic term in (\ref{gammaundergaussian}) becomes anisotropic with respect to the directions along and perpendicular to the
magnetic field due to the rotational symmetry breaking by the magnetic field. One can readily show that $a=b$ at zero magnetic field as expected. Moreover, the coefficients satisfy the inequality $a\ge b$ regardless of solutions to the gap equation(see appendix D for details).

The bosonic spectrum at low momentum is determined by the pole of $\Gamma(\omega,{\bf p})$ and reads
\begin{equation}
\omega=\omega_b\equiv\frac{p_3^2+\kappa {\bf p}_\perp^2}{4\bar{\mu}},
\end{equation}
with the anisotropicity factor
\begin{equation}
\kappa=\frac{b}{a}\le 1.
\end{equation}
In contrast to the non-relativistic case, in which the mass of the pair is $2m$, the mass is replaced by the $2\bar{\mu}$, which is magnetic field dependent. Once $\bar{\mu}$ becomes smaller than $m$, stable bosonic pairs appear and they dominate the total number density at
sufficienly low temperature. The condensation temperature is determined by setting the chemical potential at the solution of the mean field equation (\ref{regularizedgapeq}), i.e. $\mu=\bar\mu$, and the phase factor follows
\begin{equation}
\delta(\omega,{\bf q})=\pi\theta(\omega-\omega_b).\label{phasefactor}
\end{equation}
where $\theta(x)$ is the Heaviside step function. Substituting (\ref{phasefactor}) into (\ref{numberdensity}), we obtain
\begin{equation}
n=2\int \frac{d^3\bf p}{(2\pi)^3}\left[\exp\left(\frac{p_z^2+\kappa p_\perp^2}{4\bar \mu T_C}\right)-1\right]^{-1},
\label{BECcondition}
\end{equation}
where $n_0$ the fermionic contribution had been ignored under the condition $T_C<<m-\bar\mu$ of our approximation.
Solving (\ref{BECcondition}) for $T_C$, we end up with the formula (\ref{transitiontemperature}) reported in section I.

\section{Bose-Einstein Condensation in a Magnetic Field}

In this section, we shall explore the BEC critical temperature in view of (\ref{regularizedgapeq}) and (\ref{transitiontemperature}). The results display a nontrivial dependence
on the strength of the coupling constant and magnetic field.
The strong coupling and weak coupling domains correspond to $G_R>G_R^C$ and $G_R<G_R^C$ respectively. For the discussion below, it is convenient to
introduce a dimensionless critical temperature
\begin{equation}
t_C\equiv \frac{T_C}{T_C^0}=\kappa^{\frac{2}{3}}\frac{{\bar\mu_0}}{\bar\mu}
\label{dimlTc}
\end{equation}
where $T_C^0$ is given by (\ref{transitiontemperature}) with $\kappa=1$ and $\bar\mu=\bar\mu_0$. In the strong coupling domain, $\bar\mu_0$ is the solution of the gap
equation at zero magnetic field and $T_C^0$ is the critical temperature of a bound pair at zero magnetic field. In the weak coupling domain, we set $\bar\mu_0=m$
and $T_C^0$ the critical temperature of a nonrelativistic boson with mass $2m$ at zero magnetic field.
The (inverse)magnetic catalysis implies
increasing(decreasing) $t_C$ with magnetic field.

Let us consider first the weak magnetic field limit $eB<<m^2-\mu^2$, where analytic method can be applied. In the strong coupling domain, we expand the RHS of the gap
equation to the quadratic order in $(eB)^2$ with the aid of Stirling formula for the Gamma and di-Gamma functions in (\ref{regularizedgapeq}) and the Euler-Maclaurin formula
for the infinite series there and obtain
\begin{align}
\nonumber &\frac{1}{G_R}-\frac{1}{G_R^C}=\frac{\mu^2-m^2}{2\pi^2}\ln\Lambda^2+\frac{\mu^2-m^2}{\pi^2}\\
&-\frac{\mu\sqrt{m^2-\mu^2}}{\pi^2}\tan^{-1}\frac{\mu}{\sqrt{m^2-\mu^2}}+c_1(eB)^2+O((eB)^4).
\end{align}
with the solution
\begin{equation}
\bar\mu\simeq\bar\mu_0\left[1-\frac{c_1}{c_2}\frac{(eB)^2}{\bar\mu_0^2(m^2-\bar\mu_0^2)}\right],
\end{equation}
where
\begin{equation}
c_1=\frac{1}{12\pi^2}\left(1+\frac{\bar\mu_0}{\sqrt{m^2-\bar\mu_0^2}}\tan^{-1}\frac{\bar\mu_0}{\sqrt{m^2-\bar\mu_0^2}}\right),
\end{equation}
and $c_2$ is the derivative of RHS of (\ref{regularizedgapeq}) with respect to $\mu$ at $B=0$ and $\mu=\bar\mu_0$, given by
\begin{equation}
c_2=\frac{1}{\pi^2}(\ln\Lambda^2+1)+\frac{2\bar\mu_0^2-m^2}{\pi^2\bar\mu_0\sqrt{m^2-\bar\mu_0^2}}\tan^{-1}\frac{\bar\mu_0}{\sqrt{m^2-\bar\mu_0^2}}.
\end{equation}
The anisotropicity in this approximation reads
\begin{equation}
\kappa\simeq 1-\frac{d_1}{d_2}\left(\frac{eB}{m^2-\bar\mu_0^2}\right)^2,
\end{equation}
with
\begin{widetext}
\begin{align}
\nonumber d_1&=\frac{(eB)^2}{3}\int_0^1 dx\int_0^\infty d\lambda\lambda^2 x e^{\lambda[4x(1-x)\bar{\mu}_0^2-m^2]}\left\{\frac{1}{\lambda}(2+x-6x^2+3x^3)+x(1-x)^2\left[4x(1-x)\bar{\mu}_0^2+m^2\right]\right \}\\
&=\frac{3m^2-2\bar{\mu}_0^2}{32\bar{\mu}_0^2}-\frac{9m^4-28m^2\bar{\mu}_0^2+16\bar{\mu}_0^4}{96\bar{\mu}_0^3\sqrt{m^2-\bar{\mu}_0^2}}\tan^{-1}\frac{\bar{\mu}_0}{\sqrt{m^2-\bar{\mu}_0^2}},
\end{align}
\end{widetext}
and
\begin{align}
\nonumber d_2=&\sum_s C_s\int_0^1 dx\int_0^\infty d\lambda x(1-x)e^{\lambda\left[4x(1-x)\bar{\mu}_0^2-M_s^2\right]}\\
\nonumber&\times\left[\frac{3}{\lambda}+4x(1-x)\bar{\mu}_0^2+M_s^2\right] \\
=&\frac{1}{2}\ln\Lambda^2-\frac{m^2-2\bar{\mu}_0^2}{2\bar{\mu}_0\sqrt{m^2-\bar{\mu}_0^2}}\tan^{-1}\frac{\bar{\mu}_0}{\sqrt{m^2-\bar{\mu}_0^2}}+\frac{2}{3}.
\end{align}
It follows that the dimensionless critical temperature
\begin{equation}
t_C=1-\rho\left(\frac{eB}{m^2-\bar\mu_0^2}\right)^2,
\end{equation}
with the coefficient
\begin{equation}
\rho = -\frac{c_1}{c_2}\frac{m^2-\bar\mu_0^2}{\bar\mu_0^2}+\frac{2d_1}{3d_2}.
\end{equation}
The numerical values of the coefficient $\rho$ are tabulated in Table I for different couplings and cutoffs in the strong coupling domain and indicate an
inverse magnetic catalysis at a weak magnetic field.


\begin{figure}
\includegraphics[height=7cm]{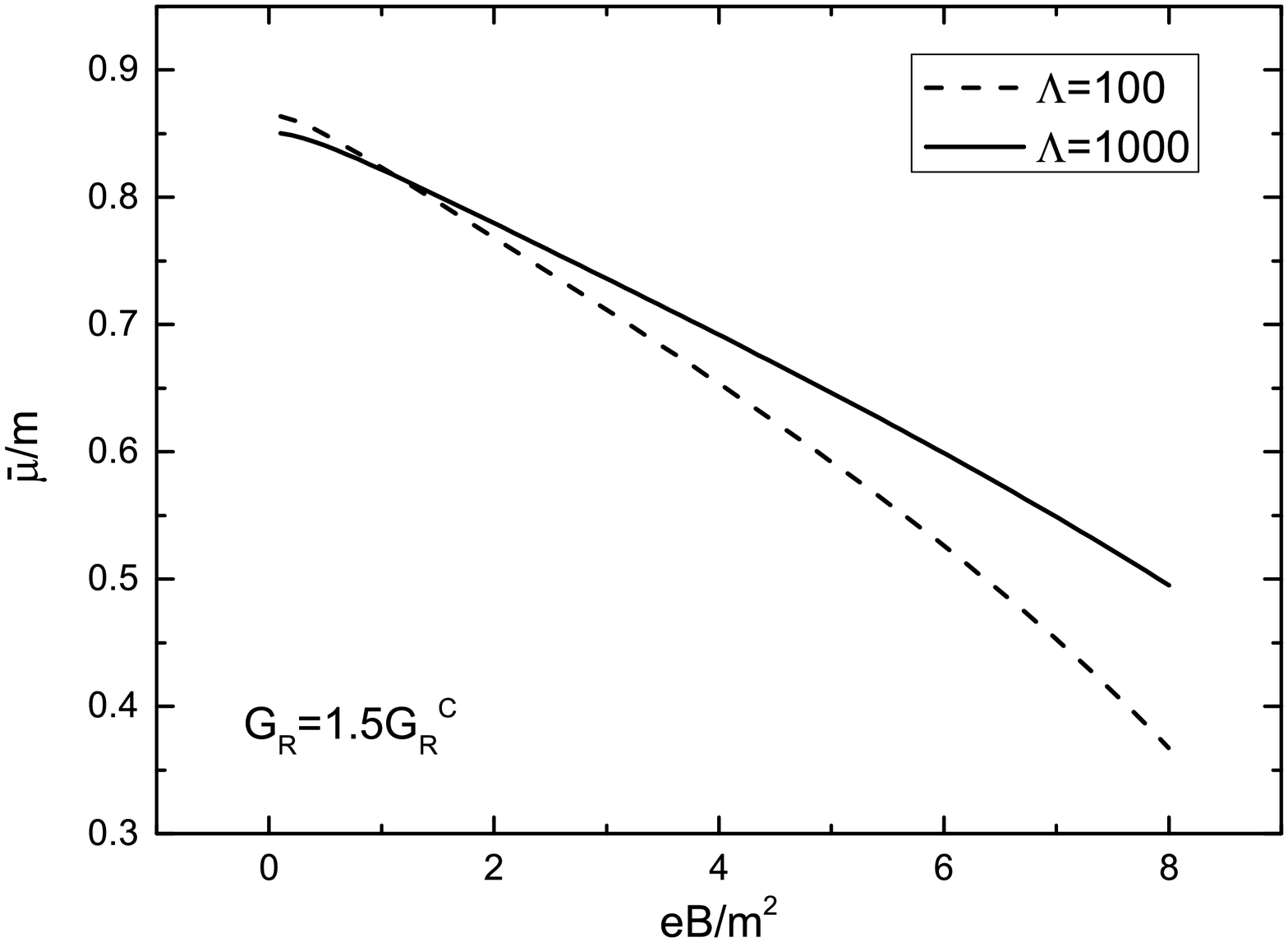}
\caption{\label{fig:epsart} The scaled bound state mass versus the dimensionless magnetic field in strong coupling domain. The solid and dashed line corresponds to an UV cutoff $\Lambda=1000$ and $\Lambda=100$, respectively.}
\end{figure}

\begin{figure}
\includegraphics[height=7cm]{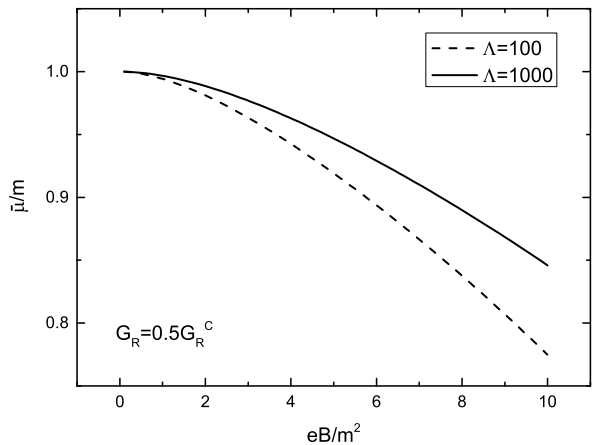}
\caption{\label{fig:epsart} The scaled bound state mass versus the dimensionless magnetic field in weak coupling domain. The solid and dashed line corresponds to an UV cutoff $\Lambda=1000$ and $\Lambda=100$, respectively.}
\end{figure}

Coming to the weak coupling domain, the bound magnetic field is the only catalyst of the bound state and a weak magnetic field triggers a weak bound state,
nonrelativistic formulas in \cite{Feng} can be carried over in the weak field limit.
\begin{equation}
\bar\mu\simeq m-\frac{1}{2}m\omega_B^2a_s^2
\end{equation}
and
\begin{equation}
\kappa\simeq 4eBa_s^2<<1
\end{equation}
with the scattering length $a_s$ extracted from
\begin{equation}
-\frac{m}{4\pi a_s}=\frac{1}{G_R}-\frac{1}{G_R^C}.
\end{equation}
The LLL approximation works here and the usual magnetic catalysis emerges (increasing with $B$). The details of the non-relativistic approximation of the gap equation can be
found in Appendix D.

\begin{table}
\caption{\label{tab:table}Numerical values of $\rho$ for different couplings and cutoffs in the strong coupling domain. }
\begin{ruledtabular}
\begin{tabular}{ccc}
 & $1.1G_R^C$ &$1.5G_R^C$\\
\hline
$\Lambda=100$& $1.863\times 10^{-2}$ & $1.159\times 10^{-2}$\\
$\Lambda=1000$&  $1.402\times 10^{-2}$ &  $7.992\times 10^{-3}$ \\
\end{tabular}
\end{ruledtabular}
\end{table}

\begin{figure}
\includegraphics[height=7cm]{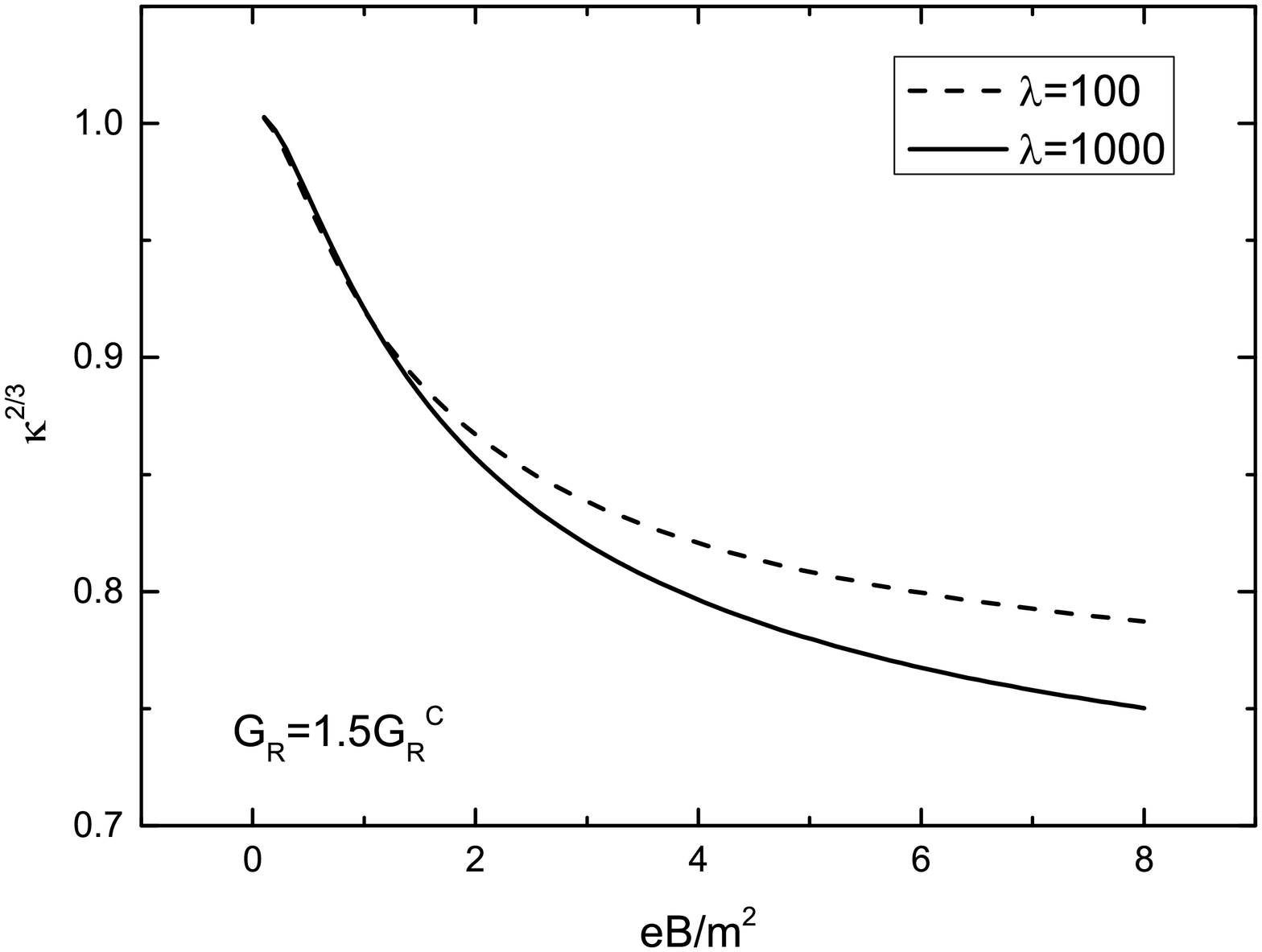}
\caption{\label{fig:epsart} The anisotropicity $\kappa$ versus the dimensionless magnetic field in strong coupling domain. The solid and dashed line corresponds to  solutions to the gap equation with an UV cutoff $\Lambda=1000$ and $\Lambda=100$, respectively.}
\end{figure}

\begin{figure}
\includegraphics[height=7cm]{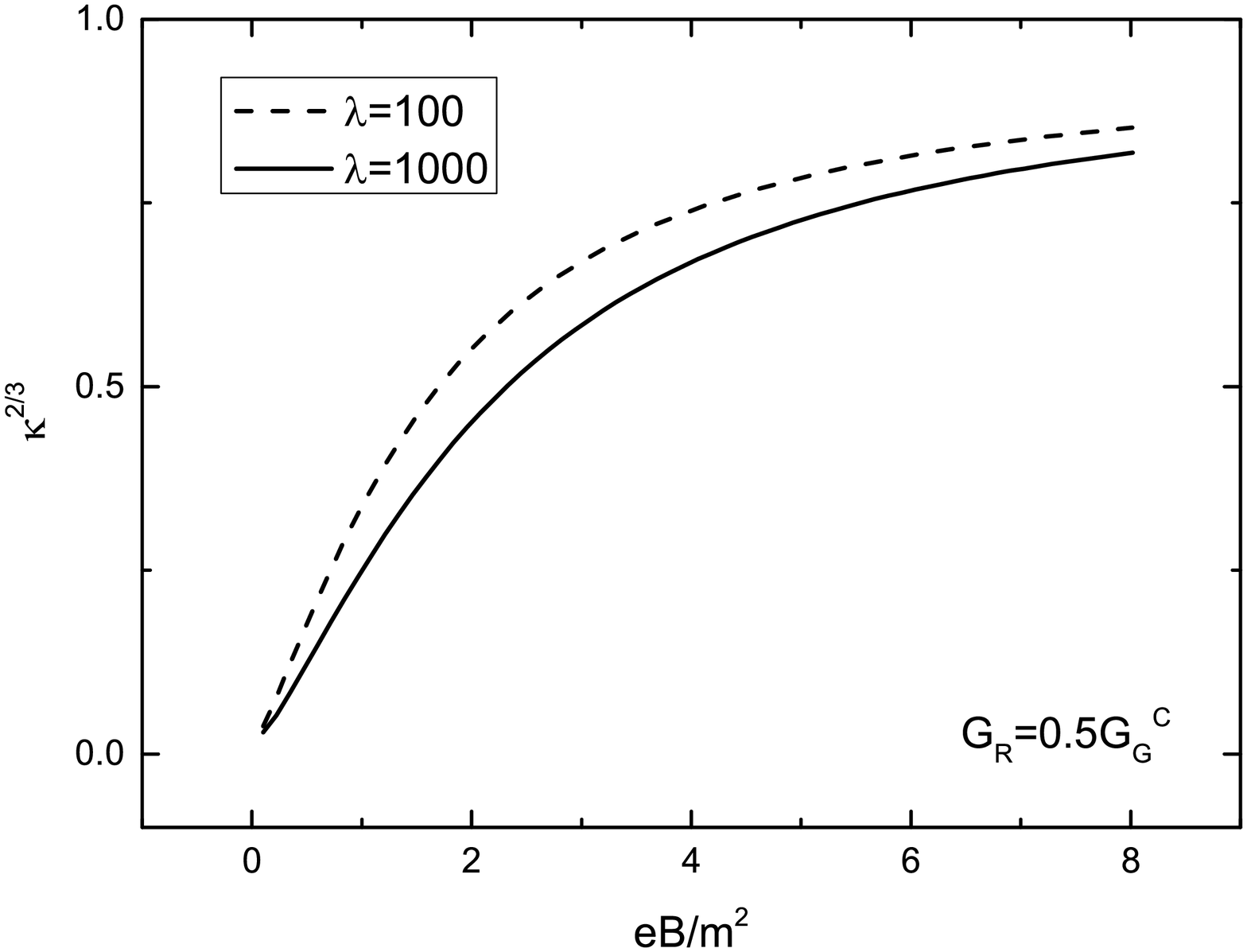}
\caption{\label{fig:epsart} The anisotropicity $\kappa$ versus the dimensionless magnetic field in weak coupling domain. The solid and dashed line corresponds to  solutions to the gap equation with an UV cutoff $\Lambda=1000$ and $\Lambda=100$, respectively.}
\end{figure}

\begin{figure}
\includegraphics[height=7cm]{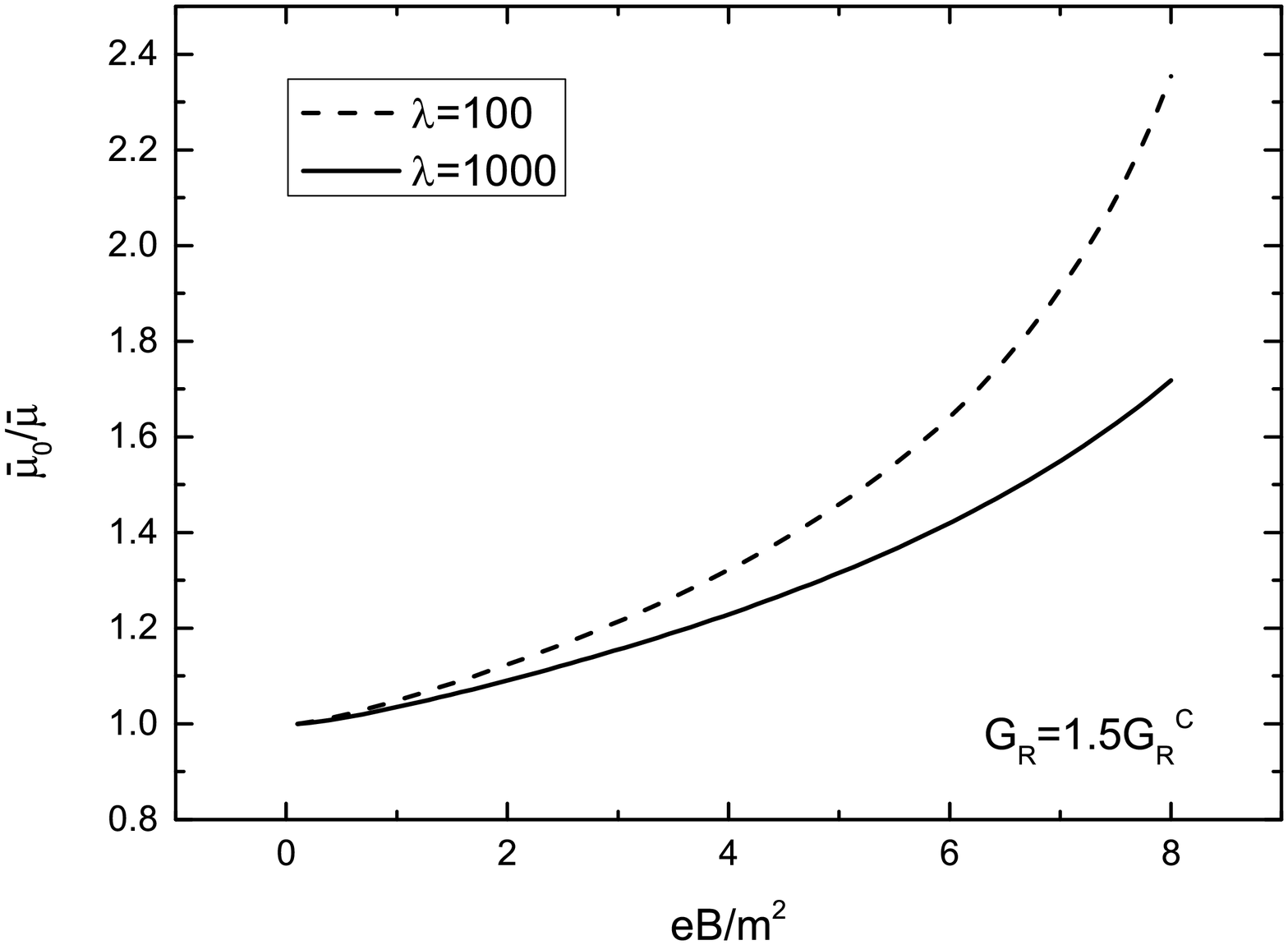}
\caption{\label{fig:epsart} The ratio $\bar{\mu}_0/\bar{\mu}$ versus the dimensionless magnetic field in strong coupling domain. The solid and dashed line corresponds to  solutions to the gap equation with an UV cutoff $\Lambda=1000$ and $\Lambda=100$, respectively.}
\end{figure}

\begin{figure}
\includegraphics[height=7cm]{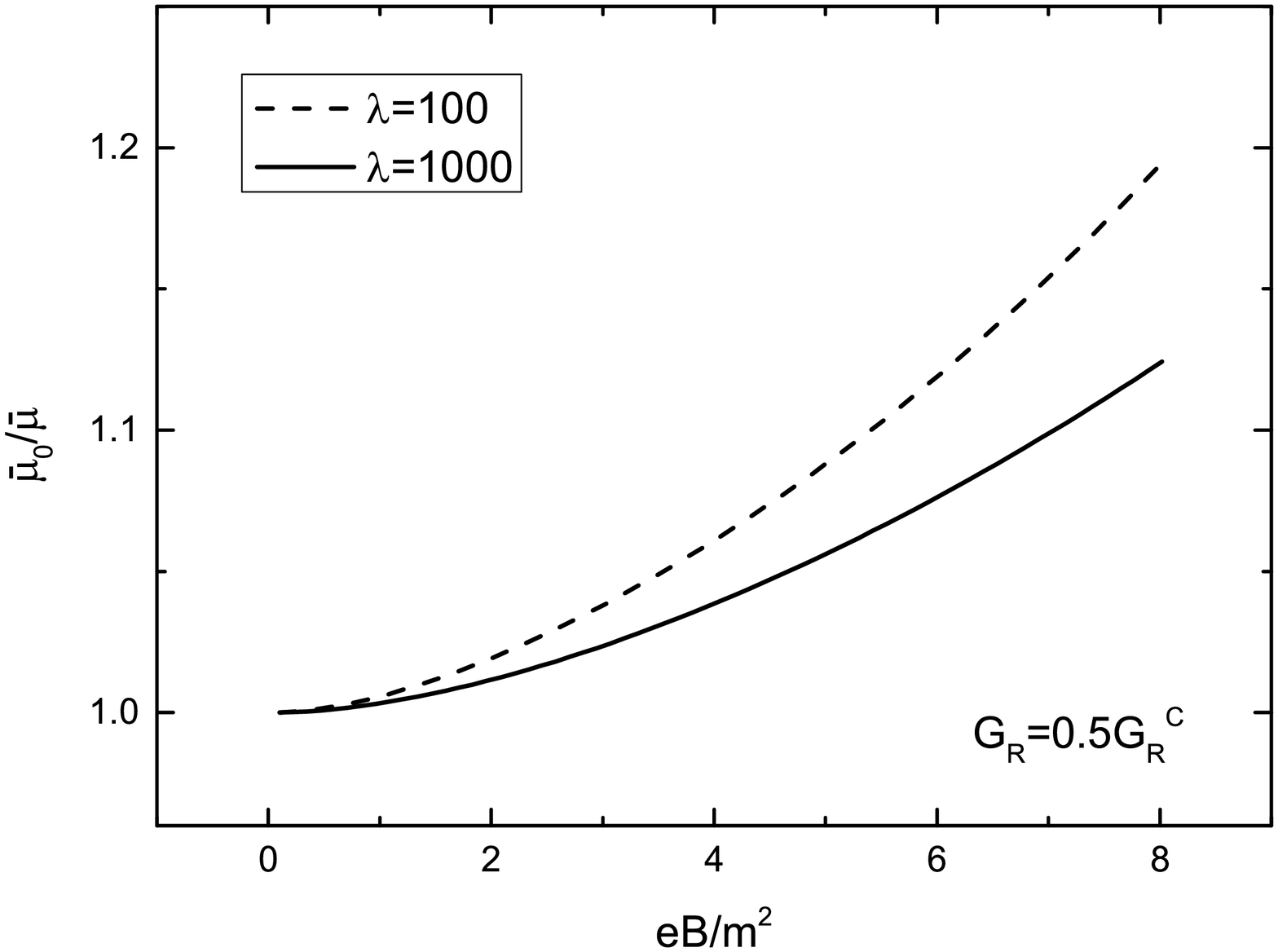}
\caption{\label{fig:epsart} The ratio $\bar{\mu}_0/\bar{\mu}$ versus the dimensionless magnetic field in weak coupling domain. The solid and dashed line corresponds to  solutions to the gap equation with an UV cutoff $\Lambda=1000$ and $\Lambda=100$, respectively.}
\end{figure}

\begin{figure}
\includegraphics[height=7cm]{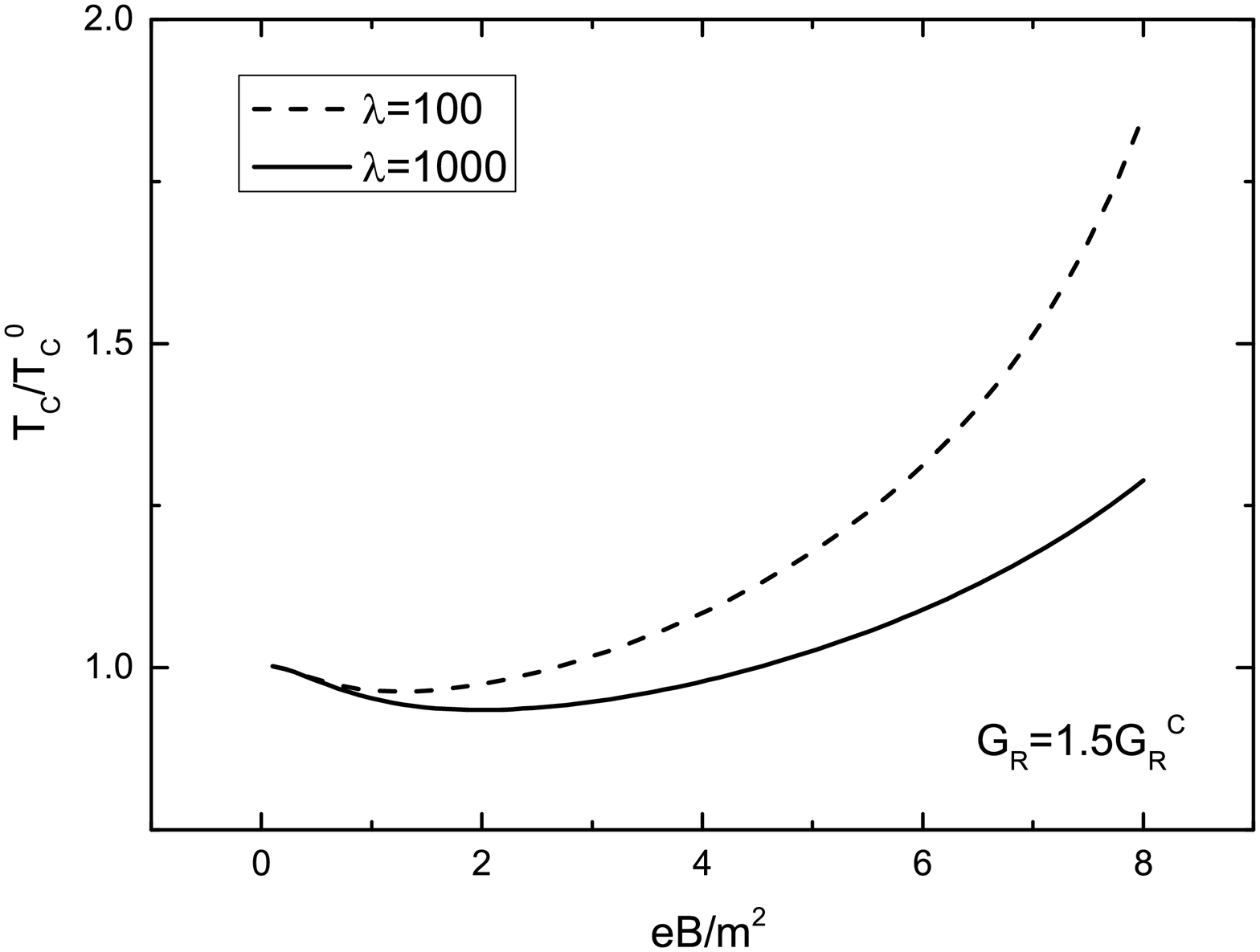}
\caption{\label{fig:epsart} The condensation temperature versus the dimensionless magnetic field in strong coupling domain. The solid and dashed line corresponds to  solutions to the gap equation with an UV cutoff $\Lambda=1000$ and $\Lambda=100$, respectively.}
\end{figure}

\begin{figure}
\includegraphics[height=7cm]{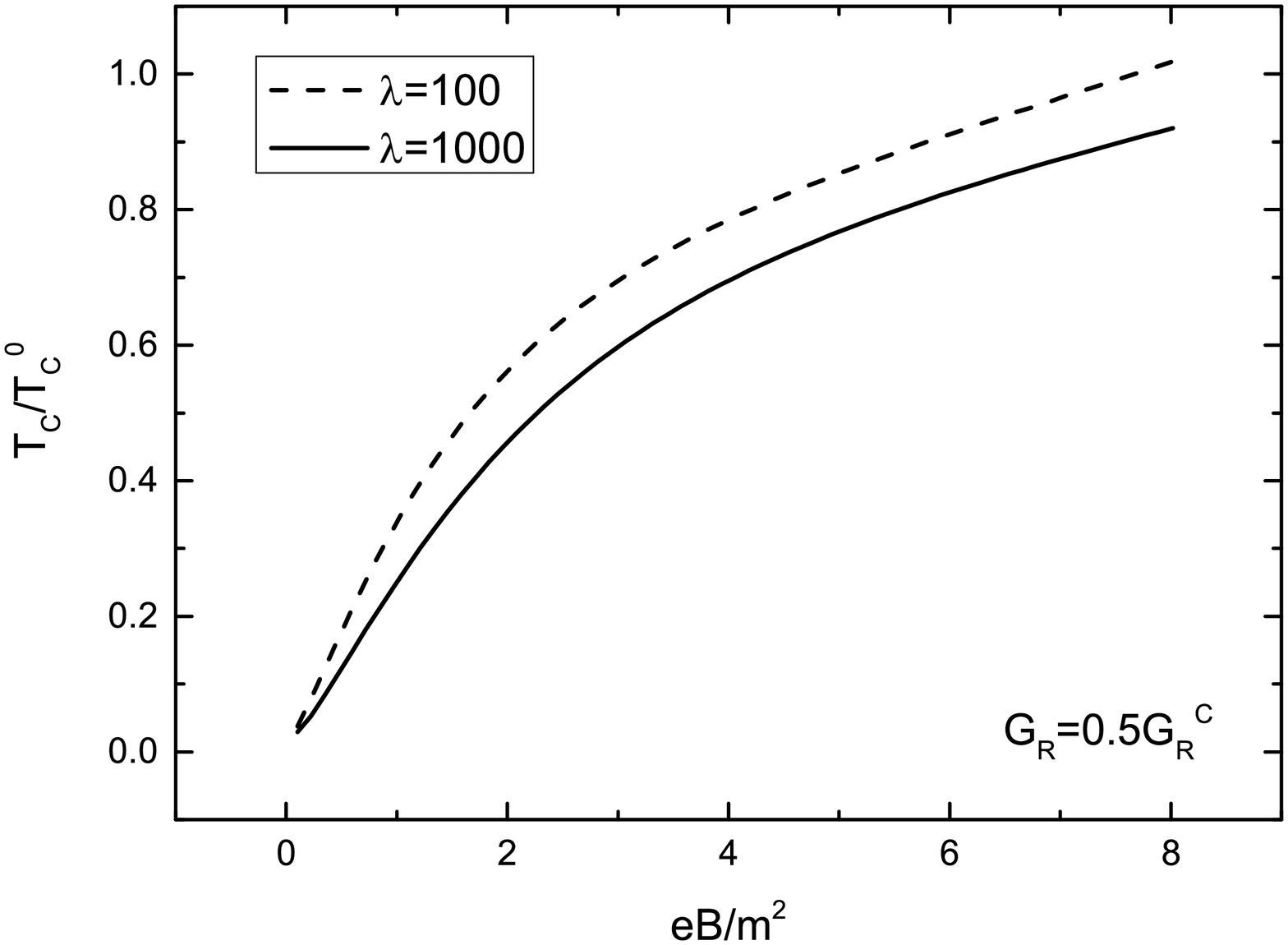}
\caption{\label{fig:epsart} The condensation temperature versus the dimensionless magnetic field in weak coupling domain. The solid and dashed line corresponds to  solutions to the gap equation with an UV cutoff $\Lambda=1000$ and $\Lambda=100$, respectively.}
\end{figure}

Beyond the weak field limit, we solve the gap equation (\ref{regularizedgapeq}) numerically at two different coupling domains that are mildly above and blow the
critical coupling with two different sets of UV cutoff and the results are presented in Fig.1 and Fig.2. In both cases, the chemical potential at the critical temperature
decreases with an increasing
magnetic field indicating an increasing binding energy of the bound pairs, consistent with the magnetic catalysis.
In strong coupling domain, however, bound pairs may appear at zero magnetic field in contrast to the case in weak coupling domain, in which the pairs can only support by
magnetic catalysis. Therefore, the curve in Fig.1 starts from $\bar\mu/m = \bar\mu_0/m < 1$ at $B=0$ while the curve in Fig.2 from $\bar\mu/m=1$ at $B=0$.
In both cases, the results display an explicit dependence on the UV cutoff, which is intrinsic in the (3+1) dimensional model with contact interaction.
But the dependence does not change the results qualitatively.
The reason for the cross between the solid and dashed curves at lower magnetic field can be explained as follows.  One can show that
\begin{equation}
\frac{d\mu}{d\Lambda}=-\frac{\left(\frac{\partial{\cal G}}{\partial\Lambda}\right)_\mu-\frac{1}{\Lambda\pi^2}\frac{m^2}{r}}{\left(\frac{\partial{\cal G}}{\partial\mu}\right)_\Lambda}=\frac{1}{\Lambda\pi^2}\left(\frac{m^2}{r}-\mu^2\right)\left[\frac{\partial{\cal G}}{\partial\mu}\right]^{-1}.\label{muoflambda}
\end{equation}
with $G_R=rG_R^C$ and ${\cal G}(\mu,\Lambda)$ the RHS of the gap equation (\ref{regularizedgapeq}). It had been proved in the appendix that $\partial{\cal G}/\partial\mu>0$ and (\ref{muoflambda}) thus changes sign at a critical value $\mu/m=\sqrt{1/r}$. One therefore expects a crossing  of the two sets of solutions corresponding to different UV cutoffs at strong coupling domain with $r>1$ as shown in Fig.1. Such crossing should be absent in weak coupling since $\mu/m\leq 1$.

The magnetic field dependence of the dimensionless critical temperature (\ref{dimlTc}) is determined by that of the anosotropicity factor $\kappa$ and the ratio
$\bar\mu/\bar\mu_0$, which are plotted in Fig.3 to Fig.8.
As we can see, the anisotropicity decreases (increases) with magnectic field in strong (weak) coupling domains while the ratio $\bar\mu_0/\bar\mu$ increases with
magnetic field in both domains. Consequently, the competetion between the two opposite field dependence of $\kappa^{\frac{2}{3}}$ and $\bar\mu_0/\bar\mu$ in the strong
coupling domain yields an inverse magnetic catalysis at lower magnetic field but magnetic catalysis at higher magnetic field. In the weak coupling domain, however,
both factors increases with the magetic field and the usual magnetic catalysis emerges.

\section{Conclusions and Outlooks}

In this paper, we have investigated the BEC of bound pairs composed by two equally and oppositely charged relativistic fermions in the presence of a magnetic field.
This is inspired by the inverse magnetic catalysis effect found in lattice QCD and is a generalization of our non-relativistic work \cite{Feng}. The purpose is to explore
the interplay between the pairing dynamics and the fluctuations of bound pairs in a magnetic field, the latter of which underlies the Mermin-Wigner-Coleman theorem on the absence of
a long range order in lower dimensions. A Nambu-Jona-Lasinio(NJL)-like pairing interaction is considered and the fluctuation
is examined within the Gaussian approximation, which is valid at a sufficiently low density. Unlike the case without a magnetic field where a bound state emerges
for the coupling strength above a certain threshold, the bound states occurs also in the weak coupling domain because of the $1D$ nature of the fermionic spectrum within
LLL. The binding energy of a bound pair is found to increase with the magnetic field as a consequence of the magnetic catalysis.

The dependence of the BEC critical temperature on the magnetic field depends on the coupling strength. In weak coupling domain, we found the usual magnetic catalysis. In strong coupling domain, unlike the nonrelativitic case, the inverse magnetic catalysis appears only in small magnetic field region and magnetic catalysis takes place again when magnetic field becomes sufficiently large. The reason for this difference is the dependence of the mass of bound pairs on the magnetic field in relativistic case. The condensation temperature is thus determined by two factors, the anisotropy and the ratio between the mass of bound pairs at zero magnetic field and the one with magnetic field dependence, both depends on the magnetic field. In strong coupling domain, the two factors have opposite field dependence and yield the inverse magnetic catalysis in small magnetic filed region but magnetic catalysis at large field. In weak coupling domain, both factors increase with the magnetic field and magnetic catalysis prevails.  This variation indicates that the enhanced fluctuations by the anisotropy in the kinetic terms of the bound pairs is less dominant as compared with the nonrelativistic case.

Although the system we have studied shares the same physics as the chiral condensate in QCD regarding general prperties of a long range order
and its fluctuations and is analytically tractable. The QCD dynamics, however, are expected to be far more complicated and other mechanisms may also contribute
the simulated magnetic field dependence of the condensation temperature. For example, the quark loop that is influenced by the magnetic field can impact on the running of
QCD coupling.
In recently works based on functional renormalization group studies\cite{Paw,Braun}, inverse magnetic catalysis at a low magnetic field and a delayed magnetic catalysis
at a high field had been found. The magnetic field at the turning point (minimum $T_c$) between decreasing and increasing $T_c$ is expected to be of order of several
$GeV^2$. AdS/QCD model suggests the same behavior and similar value of magnetic field at the turning point\cite{Mamo}. It is shown by one lattice calculation\cite{endrodi}, however,
that even at an extremely strong field $eB=3.25GeV^2$, inverse magnetic catalysis still prevails. The reconciliation is still lacking at the present stage.

In addition to the theoretical values of our present study, the physics involved
may also be applicable in cold dense quark matter, such as the color-flavor-locking superconducting phase and the planar phase in single flavor
superconductivity\cite{mcfl,FHR}. They are of interests in dense neutron stars, where magnetic field up to $10^{14}G$ to $10^{16}G$ appears
on the surface and even higher magnetic field exists in the interior.

\begin{acknowledgments}
This research is partly supported by the
Ministry of Science and Technology of China (MSTC) under the "973" Project No. 2015CB856904(4).  B. F. is supported by NSFC under grant No. 11305067, No. 11535005
and the Fundamental Research Funds for the Central Universities, 　HUST: No. 2015TS016.
D-f. Hou and H-c Ren are partly supported by NSFC under Grant Nos. 11375070, 11135011 and  11221504. P-p Wu is supported by the Exploratory fund of Henan polytechnic university, No. 13156041.

\end{acknowledgments}

\appendix

\section{}

In this appendix, the reduction of the gap equation (\ref{gapeqinappendix}) to the form (\ref{regularizedgapeq}) shall be presented. The first term inside the brackets of (\ref{gapeqinappendix}) can be evaluated as
\begin{align}
\nonumber &\sum_{l=0}^{\infty}\sum_sC_s\ln\frac{1}{2leB+M_s^2}\\
\nonumber=&\sum_sC_s\ln\frac{2eB}{M_s^2}\\
&+\sum_{l=1}^{\infty}\sum_sC_s\left[\ln\frac{1}{1+M_s^2/(2leB)}+\frac{M_s^2}{2eB}\ln\left(1+\frac{1}{l}\right)\right].
\end{align}
Using the infinite product representation of Gamma function
\begin{equation}
\Gamma(z)=\frac{1}{z}\prod_{n=1}^{\infty}\left[\left(1+\frac{z}{n}\right)^{-1}\left(1+\frac{1}{n}\right)^z\right],
\end{equation}
one has
\begin{align}
\sum_{l=0}^{\infty}\sum_sC_s\ln\frac{1}{2leB+M_s^2}
=\sum_sC_s\ln\Gamma\left(\frac{M_s^2}{2eB}\right),
\end{align}
Therefore the gap equation (\ref{gapeqinappendix}) becomes
\begin{align}
\frac{1}{G}=\frac{eB}{2\pi^2}\sum_sC_s\left[\ln\Gamma\left(\frac{M_s^2}{2eB}\right)-\frac{1}{2}\ln\frac{1}{M_s^2}+\sum_{l=0}^{\infty}\alpha_l\theta_s\tan\theta_s\right]\label{intermediate}.
\end{align}
with
\begin{equation}
\theta_s=\tan^{-1}\frac{\mu}{\sqrt{2leB+M_s^2-\mu^2}}\label{thetas},
\end{equation}
It can be further simplified for $M_s>>\mu$ and $M_s>>\sqrt{eB}$ by eliminating the terms that vanishes in the limit $M_s\rightarrow \infty$\footnote{for $M_s$ with $s>0$ since $M_0^2=m^2$.}.
For this purpose, we rewrite (\ref{intermediate}) as
\begin{align}
\nonumber\frac{1}{G}=&\frac{eB}{2\pi^2}\sum_sC_s\left\{\ln\Gamma\left(\frac{M_s^2}{2eB}\right)-\frac{1}{2}\ln\frac{1}{M_s^2}+\sum_{l=0}^{\infty}\alpha_l\tan^2\theta_s\right.\\
&+\left.
\sum_{l=0}^{\infty}\alpha_l\left[\theta_s\tan\theta_s-\tan^2\theta_s\right]\right\}.
\end{align}
We have
\begin{align}
\nonumber&\sum_sC_s\sum_{l=0}^{\infty}\tan^2\theta_s\\
\nonumber =&\frac{\mu^2}{2eB}\sum_sC_s\left(\sum_{l=0}^{\infty}\frac{1}{l+(M_s^2-\mu^2)/(2eB)}-\sum_{l=1}^{\infty}\frac{1}{l}+\gamma\right)\\
=&-\sum_sC_s\frac{\mu^2}{2eB}\psi\left(\frac{M_s^2-\mu^2}{2eB}\right).
\end{align}
and the other infinite series converges for each $s$, which vanishes in the limit $M_s\to\infty$ for $s>0$ and can be ignored.
Applying the Stirling formulas
\begin{align}
\nonumber\ln\Gamma(z)=&\left(z-\frac{1}{2}\right)\ln z-z+\frac{1}{2}\ln(2\pi)\\
&+\sum_{r=1}^{n}\frac{(-1)^{r-1}B_r}{2r(2r-1)}z^{-2r+1}+O(z^{-2n-1}),
\end{align}
and
\begin{equation}
\psi(z)=\ln z-\frac{1}{2z}-\sum_{r=1}^{n}\frac{(-1)^{r-1}B_r}{2r}z^{-2r}+O(z^{-2n-2}),
\end{equation}
with $B_r$ the Bernoulli numbers, $B_1=1/6, B_2=1/30,...$. We thus have, for $s>0$
\begin{align}
\nonumber\ln\Gamma\left(\frac{M_s^2}{2eB}\right)\cong &\left(\frac{M_s^2}{2eB}-\frac{1}{2}\right)\ln\frac{M_s^2}{2eB}-\frac{M_s^2}{2eB}+\frac{1}{2}\ln(2\pi)\\
&+O(z^{-1}).
\end{align}
and
\begin{equation}
\psi\left(\frac{M_s^2-\mu^2}{2eB}\right)\cong\ln\frac{M_s^2}{2eB}+O(z^{-1}).
\end{equation}
We end up with the form of (\ref{regularizedgapeq}) of the gap equation.

\section{}
In this appendix, we shall provide analytical proofs of the monotinic dependence of the bound state mass and the anisotropy
factors on the magnetic field. Because of the complication introduced by Pauli-Villars regulators, we limit our scope within
the approximaton that maintains only the logarithmic terms in the regulator masses.

Denoting the RHS of the gap equation (\ref{gapeqinappendix}) by ${\cal G}(\mu,B)$, we have
\textit{Lemma}: (1) For a given $B$, ${\cal G}$ is a monotonically increasing function of $\mu$ for $0<\mu<m$;
         (2) For a given $\mu$ within $(0,m)$, ${\cal G}$ is a monotonically increasing function of $B$.

Proof: (1) The partial derivative of ${\cal G}$ with respect to $\mu$ reads:
\begin{align}
\nonumber\left(\frac{\partial {\cal G}}{\partial\mu}\right)_B
=&\frac{eB}{2\pi^2}\sum_{l=0}^\infty\alpha_l\sum_sC_s{\frac{2leB+M_s^2}{(2leB+M_s^2-\mu^2)^{\frac{3}{2}}}}\\
&\times\left(\theta_s+\frac{1}{2}\sin2\theta_s\right),
\end{align}
with $\theta_s$ defined in (\ref{thetas}) and
\begin{equation}
\sin2\theta_s=\frac{2\tan\theta_s}{1+\tan^2\theta_s}=\frac{2\mu\sqrt{2leB+M_s^2-\mu^2}}{2leB+M_s^2},
\end{equation}
and $\theta_s\ge\frac{1}{2}\sin2\theta_s$.
Let us divide the above derivative into two terms, $D_1$ and $D_2$, with $D_2$ carrying the logarithmic UV divergence and $D_1$
not. We have
\begin{align}
\nonumber D_1=&\frac{eB}{2\pi^2}\sum_{l=0}^\infty \alpha_l\sum_sC_s\left[\frac{2leB+M_s^2}{(2leB+M_s^2-\mu^2)^{\frac{3}{2}}}\right.\\
\nonumber&\times\left.\left(\theta_s+\frac{1}{2}\sin2\theta_s\right)-\frac{2\mu}{2leB+M_s^2}\right]\\
\simeq&\frac{eB\mu}{\pi^2}\left[\sum_{l=0}^\infty \alpha_l\left(\frac{1}{2leB+m^2-\mu^2}-\frac{1}{2leB+m^2}\right)\right]\ge 0,
\end{align}
and
\begin{equation}
D_2=\frac{\mu}{\pi^2}\left[\ln\frac{M^2}{m^2}-\psi\left(\frac{m^2}{2eB}\right)-\frac{eB}{m^2}+\ln\frac{m^2}{2eB}\right].
\end{equation}
where we have assumed that the terms with negative power of the regulator masses dropped from $D_1$ are too small to offset the direction
of the inequality. It follows from the Binet formula
\begin{equation}
\ln\Gamma(z)=z(\ln z-1)+\ln\frac{2\pi}{z}+2\int_0^\infty dt\frac{\tan^{-1}(t/z)}{e^{2\pi t}-1},
\label{B1}
\end{equation}
and its derivative
\begin{equation}
\psi(z)=\ln z-\frac{1}{2z}-2\int_0^\infty dt\frac{1}{(t^2+z^2)(e^{2\pi t}-1)},
\label{B2}
\end{equation}
that $D_2\ge 0$. Consequently
\begin{equation}
\left(\frac{\partial {\cal G}}{\partial\mu}\right)_B\ge 0.
\end{equation}


(2) Taking the derivative of (\ref{gapeqinappendix}) with respect to $B$, we find that
\begin{equation}
\left(\frac{\partial {\cal G}}{\partial B}\right)_\mu=E_1+E_2.
\end{equation}
where
\begin{align}
\nonumber E_1=&\frac{e\mu}{2\pi^2}\sum_{l=0}^\infty\alpha_l\sum_sC_s\left[\frac{leB+M_s^2-\mu^2}{(2leB+M_s^2-\mu^2)^{\frac{3}{2}}}
\theta_s\right.\\
\nonumber&-\left.\frac{leB\mu}{(2leB+M_s^2)(2leB+M_s^2-\mu^2)}\right]\\
\nonumber\ge&\frac{e\mu}{4\pi^2}\sum_{l=0}^\infty\alpha_l\left[\frac{leB+m^2-\mu^2}{(2leB+m^2-\mu^2)^{\frac{3}{2}}}
\sin2\theta_0\right.\\
\nonumber&-\left.\frac{leB\mu}{(2leB+m^2)(2leB+m^2-\mu^2)}\right]\\
=&\frac{e\mu^2}{2\pi^2}\sum_{l=0}^\infty\alpha_l\frac{m^2-\mu^2}{(2leB+m^2)}\ge 0,
\end{align}
and
\begin{align}
E_2 =& \frac{1}{(2\pi)^2}\frac{\partial}{\partial B}\left[eB\sum_{l=0}^\infty\alpha_l\sum_sC_s\ln\frac{1}{2leB+M_s^2}\right]\nonumber\\
 =&\frac{1}{2\pi^2}\frac{\partial}{\partial B}\left\{eB\left[\ln\Gamma\left(\frac{m^2}{2eB}\right)-\frac{m^2}{2eB}\left(\ln \frac{m^2}{2eB}-1\right)\right.\right.\nonumber\\
&-\left.\left.\frac{1}{2}\ln\frac{4\pi eB}{m^2}\right]\right\}\ge 0.
\end{align}
with the inequality following from (\ref{B1}) and (\ref{B2}). The lemma is proved.
For the solution of the gap equation,
\begin{equation}
\frac{d\mu}{dB}=-\frac{\left(\frac{\partial {\cal G}}{\partial B}\right)_\mu}{\left(\frac{\partial {\cal G}}{\partial \mu}\right)_B}
\le 0.
\end{equation}
and we arrive at
\textit{Theorem}: The bound state mass is a decreasing function of the external magnetic field.

\section{}

In this appendix, we shall prove analytically that the anisotropy coefficients satisfy
$a>b$. Introducing that $s=\lambda x$ and $s'=\lambda (1-x)$, we have
\begin{align}
\nonumber a-b=&\frac{2\bar{\mu}}{\pi^2}\int_0^\infty ds\int_0^\infty ds'e^{-(s'+s)m^2
+\frac{4\bar{\mu}^2 s' s}{ s'+s}}\\
&\times \frac{F(s',s)}{(s'+s)\sinh((s'+s)eB)},
\end{align}
where
\begin{widetext}
\begin{align}
\nonumber F(s',s) =&\left[\frac{2}{s'+s}+\frac{4s's\bar{\mu}^2}{s'+s}+m^2\right]
\frac{eBs's\cosh((s'+s)eB)}{s'+s}+\frac{(eB)^2s's}{(s'+s)\sinh((s'+s)eB)}
-\left[\frac{1}{s'+s}+\frac{4s's\bar{\mu}^2}{s'+s}+m^2\right]\\
\nonumber &\times\frac{\tanh (s'eB)\tanh (s eB)}{\tanh (s'eB)+\tanh (s eB)}
-\frac{2eB\tanh (s'eB)\tanh (s eB)}{\sinh((s'+s)eB)(\tanh (s'eB)+\tanh (s eB))}\nonumber\\
\nonumber =& \left[\frac{1}{s'+s}+\frac{4s's\bar{\mu}^2}{s'+s}+m^2\right]
\left[\frac{eBs's}{s'+s}\cosh((s'+s)eB)-\frac{\tanh (s'eB)\tanh (s eB)}{\tanh (s'eB)+\tanh (s eB)}\right]
+\frac{eBs's}{(s'+s)^2}\cosh((s'+s)eB)\\
\nonumber &+\frac{(eB)^2s's}{(s'+s)\sinh((s'+s)eB)}
-\frac{2eB\tanh (s'eB)\tanh (s eB)}{\sinh((s'+s)eB)(\tanh (s'eB)+\tanh (s eB))}\nonumber\\
\geq& \frac{eB s's}{s'+s}\left[\frac{1}{s'+s}+\frac{4s's\bar{\mu}^2}{s'+s}+m^2\right]
\left[\cosh((s'+s)eB)-1\right]
+\frac{eBs's}{(s'+s)^2}\left[\cosh((s'+s)eB)-\frac{(s'+s)eB}{\sinh((s'+s)eB)}\right]\nonumber
\\
\geq& \frac{eBs's}{(s'+s)^2\sinh((s'+s)eB)}\left[\sinh (2(s'+s)eB)-2(s'+s)eB\right]
\geq 0.
\end{align}
\end{widetext}
A crucial inequality employed above is that
\begin{equation}
\frac{\tanh u\tanh v}{\tanh u+\tanh v}\le\frac{uv}{u+v}.
\end{equation}
for positive $u$ and $v$. To prove it, we notice that the function $f(x)=\frac{cx}{c+x}$ is monotonically increasing for $c\ge 0$ and $x\ge 0$. Setting
$c=\tanh u$, it follows from the inequality $\tanh v\le v$ that
\begin{equation}
\frac{\tanh u\tanh v}{\tanh u+\tanh v}\le\frac{v\tanh u}{\tanh u+v},
\end{equation}
Repeating the same argument with $c=v$, we end up with
\begin{equation}
\frac{v\tanh u}{\tanh u+v}\le\frac{uv}{u+v}.
\end{equation}
and the inequality is proved.

\section{}

In this section, we shall take the nonrelativistic (NR) approximation of the gap equation (\ref{regularizedgapeq}). The conditions for the NR approximation are
$b<<m$, $\omega_B<<m$ with the NR binding energy $b\equiv(m^2-\mu^2)/(2m)$ and cyclotron frequency $\omega_B\equiv eB/m$. Let us break the infinite series on RHS
of (\ref{regularizedgapeq}) into two pieces,
\begin{equation}
S_1=\frac{eB}{\pi^2}\sum_{l=0}^{N-1}\left[\theta_0\tan\theta_0-\frac{\mu^2}{2leB+m^2-\mu^2}\right],
\end{equation}
and
\begin{equation}
S_2=\frac{eB}{\pi^2}\sum_{l=N}^{\infty}\left[\theta_0\tan\theta_0-\frac{\mu^2}{2leB+m^2-\mu^2}\right].
\end{equation}
with $\theta_s$ defined in (\ref{thetas}) and $1<<N<<\frac{m^2}{eB}$. Consequently
\begin{align}
\nonumber S_1\simeq &\frac{m^{\frac{3}{2}}\omega_B^{\frac{1}{2}}}{2\sqrt{2}\pi}\sum_{l=0}^{N-1}\frac{1}{\sqrt{l+\frac{b}{\omega_B}}}
+\frac{\mu^2}{2\pi^2}\left[\psi\left(\frac{b}{\omega_B}\right)\right.\\
&\left.-\psi\left(N+\frac{b}{\omega_B}\right)\right],
\end{align}
and the summation in $S_2$ can be approximated by an integral, i.e.
\begin{equation}
S_2\simeq\frac{1}{2\pi^2}\left(2\mu^2-\sqrt{2}\pi m^{\frac{3}{2}}\omega_B^{\frac{1}{2}}\sqrt{N+\frac{b}{\omega_B}}+\mu^2\ln\frac{2NeB}{m^2}\right).
\end{equation}
Using the formula
\begin{equation}
(l+a)^{s-1}=-\frac{\Gamma(s)}{2\pi i}\oint_C dt(-t)^{-s}e^{-(l+a)t}.
\end{equation}
with $C$ a contour on $t$-plane going around the positive real axis counterclockwisely, we find
\begin{eqnarray}
\sum_{l=0}^{N-1}\frac{1}{\sqrt{l+\frac{b}{\omega_B}}} &=& -\frac{1}{2i\sqrt{\pi}}\oint_C dt(-t)^{-\frac{1}{2}}e^{-\frac{b}{\omega_B}t}\frac{1-e^{-Nt}}{1-e^{-t}}\nonumber\\
&=& \zeta\left(\frac{1}{2},\frac{b}{\omega_B}\right)-\zeta\left(\frac{1}{2},N+\frac{b}{\omega_B}\right).
\end{eqnarray}
It follows from the Hermite formula and Stirling formula for $N>>1$ that
\begin{equation}
\zeta\left(\frac{1}{2},N+\frac{b}{\omega_B}\right)\simeq -2\sqrt{N+\frac{b}{\omega_B}},
\end{equation}
and
\begin{equation}
\psi\left(N+\frac{b}{\omega_B}\right)\simeq \ln N,
\end{equation}
Consequently
\begin{align}
\nonumber S_1+S_2\simeq & \frac{m^{\frac{3}{2}}\omega_B^{\frac{1}{2}}}{2\sqrt{2}\pi}\zeta\left(\frac{1}{2},\frac{b}{\omega_B}\right)
+\frac{\mu^2}{2\pi^2}\left[\psi\left(\frac{b}{\omega_B}\right)\right.\\
&+2\left.+\ln\frac{2eB}{m^2}\right].
\label{series}
\end{align}
Applying the Stirling formula to the Gamma function on RHS of (\ref{regularizedgapeq}) and combining with (\ref{series}), we end up with
\begin{equation}
\frac{1}{G}\simeq \frac{1}{G_c}+\frac{m^{\frac{3}{2}}\omega_B^{\frac{1}{2}}}{2\sqrt{2}\pi}
\left[\zeta\left(\frac{1}{2},\frac{b}{\omega_B}\right)-\frac{1}{2}\sqrt{\frac{\omega_B}{b}}\right].
\end{equation}
which, upon identifying $\frac{1}{G}-\frac{1}{G_c}$ with $\frac{m}{4\pi a_s}$, is the nonrelativistic gap equation obtained in \cite{Feng}.

To take the nonrelativistic limit of the coefficients $a$ and $b$ given in eqs.(\ref{coeffa}) and (\ref{coeffb}), we introduce
\begin{widetext}
\begin{align}
\nonumber  A(s)=&\frac{eB{\bar{\mu}}}{\pi^2}\int_0^1 dx\int_0^\infty d\lambda \lambda(\lambda eB)^{s-\frac{1}{2}} x(1-x)e^{\lambda\left[4x(1-x)\bar{\mu}^2-m^2\right]}\left\{\left[\frac{2}{\lambda}+4x(1-x)\bar{\mu}^2+m^2\right]\coth(\lambda eB)+\frac{eB}{\sinh^2(\lambda eB)} \right\}\\
=&\frac{{\bar{\mu}}}{\pi^2eB}\int_0^1 dx\int_0^\infty d\xi \xi^{s+\frac{1}{2}} x(1-x)e^{-\frac{m}{\omega_B}\xi(2x-1)^2+8\frac{b}{\omega_B}x(1-x)}\left\{\left[\frac{2eB}{\xi}+4x(1-x)\bar{\mu}^2+m^2\right]\coth(\xi)+\frac{eB}{\sinh^2(\xi)} \right\},
\end{align}
\end{widetext}
and
\begin{widetext}
\begin{align}
\nonumber  B(s)=&\frac{{\bar{\mu}}}{\pi^2}\int_0^1 dx\int_0^\infty d\lambda (\lambda eB)^{s-\frac{1}{2}}e^{\lambda\left[4x(1-x)\bar{\mu}^2-m^2\right]}\frac{\tanh(\lambda xeB)\tanh(\lambda(1-x)eB)}{\tanh(\lambda xeB)+\tanh(\lambda(1-x)eB)}\\
\nonumber &\times\left\{\left[\frac{1}{\lambda}+4x(1-x)\bar{\mu}^2+m^2\right]\coth(\lambda eB)+\frac{2eB}{\sinh^2(\lambda eB)}\right\}\\
\nonumber =&\frac{{\bar{\mu}}}{\pi^2}\int_0^1 dx\int_0^\infty d\xi \xi^{s-\frac{1}{2}} e^{-\frac{m}{\omega_B}\xi(2x-1)^2+8\frac{b}{\omega_B}x(1-x)}
\frac{\tanh(\xi x)\tanh(\xi(1-x))}{\tanh(\xi x)+\tanh(\xi(1-x))}\\ &\times\left\{\left[\frac{eB}{\xi}+4x(1-x)\bar{\mu}^2+m^2\right]\coth(\xi)+\frac{2eB}{\sinh^2(\xi)}\right\}.
\end{align}
\end{widetext}
and define $a$ and $b$ as their analytic continuations at $s=\frac{1}{2}$, i.e. $a=A\left(\frac{1}{2}\right)$ and $b=B\left(\frac{1}{2}\right)$.
Both integrals are convergent for ${\rm Re}s>\frac{1}{2}$ and the main contribution to the integral comes from $\xi=O(1)$. In the nonrelativistic limit, $\omega_B<<m$ and
$b<<m$, the integrand is peaked at $x=\frac{1}{2}$ and only the 2nd and 3rd terms inside the brackets dominate. It follows that
\begin{align}
\nonumber A(s) \simeq& \frac{m^3}{2\pi^2eB}\int_0^\infty d\xi\xi^{s+\frac{1}{2}}e^{-\frac{2b}{\omega_B}\xi}\coth\xi\\
\nonumber&\times\int_{-\infty}^\infty dxe^{-\frac{m}{\omega_B}\xi(2x-1)^2}
\\  =& \frac{1}{4}\left(\frac{m}{\pi}\right)^{\frac{3}{2}}\frac{1}{\sqrt{\omega_B}}\int_0^\infty d\xi\xi^se^{-\frac{2b}{\omega_B}\xi}\coth\xi,
\end{align}
Upon writing $\coth\xi=\frac{2}{1-e^{-2\xi}}+1$ and using the integral representation of the Hurwitz zeta function, we find that
\begin{equation}
a=A\left(\frac{1}{2}\right)=\frac{m^{\frac{3}{2}}}{8\pi\sqrt{\omega_B}}\left[\zeta\left(\frac{3}{2},\frac{b}{\omega}\right)
-\frac{1}{2}\left(\frac{b}{\omega_B}\right)^{-\frac{3}{2}}\right].
\end{equation}
Under the same approximation, we have
\begin{align}
\nonumber B(s)\simeq &\frac{m^3}{\pi^2eB}\int_0^\infty d\xi \xi^{s-\frac{1}{2}}
e^{-\frac{2b}{\omega_B}\xi}\coth\xi\tanh\frac{\xi}{2}\\
&\times\int_{-\infty}^\infty dx e^{-\frac{m}{\omega_B}\xi(2x-1)^2},
\end{align}
Using the identity
\begin{equation}
\coth\xi\tanh\frac{\xi}{2}=(1-e^{-\xi})^2\left[-\frac{d}{d\xi}\frac{1}{1-e^{2\xi}}+\frac{1}{1-e^{-2\xi}}\right],
\end{equation}
and integration by part, we find
\begin{widetext}
\begin{equation}
b=B\left(\frac{1}{2}\right)=\frac{2m^{\frac{3}{2}}}{\pi\sqrt{2\omega_B}}
\left\{\zeta\left(-\frac{1}{2},1+\frac{b}{\omega_B}\right)+\zeta\left(-\frac{1}{2},\frac{1}{2}+\frac{b}{\omega_B}\right)
-\frac{b}{\omega_B}\left[\zeta\left(\frac{1}{2},1+\frac{b}{\omega_B}\right)+\zeta\left(\frac{1}{2},\frac{1}{2}+\frac{b}{\omega_B}\right)\right]
+\frac{1}{4}\sqrt{\frac{\omega_B}{b}}\right\}.
\end{equation}
\end{widetext}
after some algebra.

\newpage 

\end{document}